\RequirePackage{etoolbox}

\documentclass[ba]{imsart}
\pubyear{0000}
\volume{00}
\issue{0}
\doi{0000}
\firstpage{1}
\lastpage{1}

\usepackage{amsthm}
\usepackage{amsmath}
\usepackage{natbib}
\usepackage{tikz}
\usetikzlibrary{fit,positioning,calc}
\usepackage[colorlinks,citecolor=blue,urlcolor=blue,filecolor=blue,backref=page]{hyperref}
\usepackage{graphicx}
\usepackage{amssymb}

\startlocaldefs
\numberwithin{equation}{section}
\theoremstyle{plain}

\newtheorem{prop}{Proposition}[section]
\endlocaldefs

\def\T{{ \mathrm{\scriptscriptstyle T} }}

\begin{document}

\begin{frontmatter}
\title{Bayesian Inference and Testing of \\ Group Differences in Brain Networks\thanksref{T1}}
\runtitle{Bayesian Inference on Group Differences in Brain Networks}

\begin{aug}
\author{\fnms{Daniele} \snm{Durante}\thanksref{m1}\ead[label=e1]{durante@stat.unipd.it}}
\and
\author{\fnms{David B.} \snm{Dunson}\thanksref{m2}\ead[label=e2]{dunson@duke.edu}}

\runauthor{Durante and Dunson}


\address{\thanksmark{m1} University of Padova, Department of Statistical Sciences.
Via Cesare Battisti, 241, 35121 Padova, Italy.
\printead{e1}}

\address{\thanksmark{m2} Duke University,
Department of Statistical Science.
Box 9025, Durham, NC 27708-0251 USA
\printead{e2}}
\end{aug}

\begin{abstract}
Network data are increasingly collected along with other variables of interest.  Our motivation is drawn from neurophysiology studies measuring brain connectivity networks for a sample of individuals along with their membership to a low or high creative reasoning group. It is of paramount importance to develop statistical methods for testing of global and local changes in the structural interconnections among brain regions across groups. We develop a general Bayesian procedure for inference and testing of group differences in the network structure, which relies on a nonparametric representation for the conditional probability mass function associated with a network-valued random variable.  By leveraging a mixture of low-rank factorizations,  we allow simple global and local hypothesis testing adjusting for multiplicity.  An efficient Gibbs sampler is defined for posterior computation. We provide theoretical results on the flexibility of the model and assess testing performance in simulations.  The approach is applied to provide novel insights on the relationships between human brain networks and creativity.
\end{abstract}


\begin{keyword}
\kwd{Brain Network}
\kwd{Mixture model}
\kwd{Multiple testing}
\kwd{Nonparametric Bayes}
\end{keyword}

\end{frontmatter}
\section{Introduction}
\label{sec:intro}
There has been an increasing focus on using neuroimaging technologies to better understand the neural pathways underlying human behavior, abilities and neuropsychiatric diseases. The primary emphasis has been on relating the level of activity in brain regions to phenotypes. Activity measures are available via electroencephalography (EEG) and functional magnetic resonance imaging (fMRI) --- among others --- and the aim is to produce a spatial map of the locations in the brain across which activity levels display evidence of change with the phenotype \citep[e.g][]{Gen_2002,tans_2014}.


\begin{figure}[t]
\includegraphics[width=12.3cm]{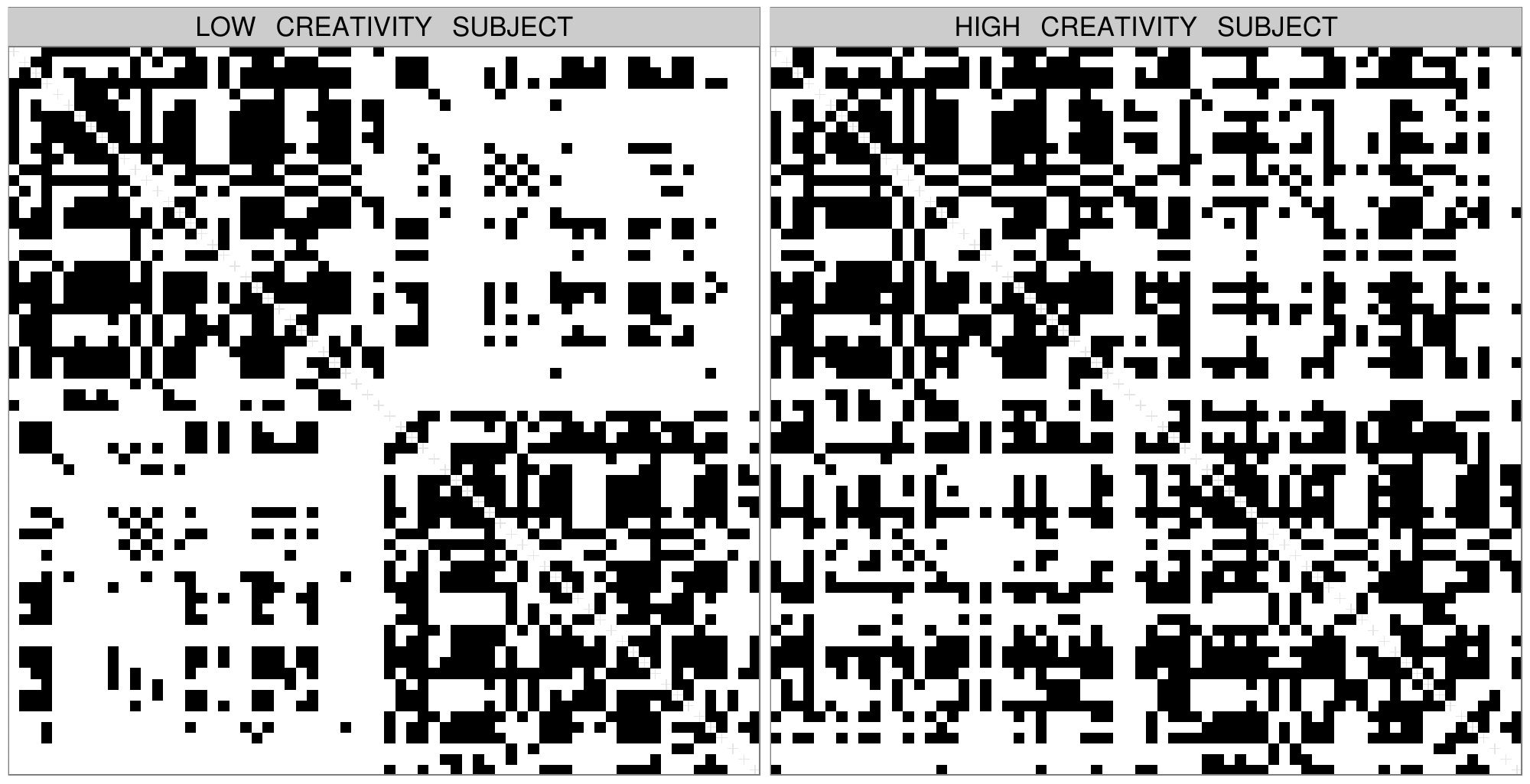}
\caption{{\small{Adjacency matrices $\boldsymbol{A}_i$ representing the brain network for two subjects in the low and high creativity group, respectively. Black refers to an edge and white to a non-edge.}}}
\label{F1}
\end{figure}

Although the above analyses remain an active area of research, more recently there has been a paradigm shift in neuroscience away from the modular approach and towards studying brain connectivity networks and their relationship with phenotypes \citep{fust_2000,fust_2006}. It has been increasingly realized that it is naive to study region-specific activity in isolation, and the overall circuit structure across the brain is a more important predictor of phenotypes \citep{bres_2010}. Brain connectivity data are now available to facilitate this task, with non-invasive imaging technologies providing accurate brain network data at increasing spatial resolution;  see \citet{stir_2008}, \citet{cra_2013} and \citet{wang_2014} for an overview and recent developments. 

A common approach for constructing brain network data is based on the covariance in activity across brain regions estimated from fMRI data. For example, one can create a functional connectivity network from the inverse covariance matrix, with low values of the precision matrix suggesting evidence of conditional independence between pairs of brain regions \citep[e.g][]{rams_2010,smit_2011,simps_2013}. Although functional connectivity networks are of fundamental interest, the recent developments in diffusion tensor imaging (DTI) technologies \citep{cra_2013} have motivated an increasing  focus on structural brain network data measuring anatomical connections made by axonal pathways. 

DTI maps the diffusion of water molecules across biological tissues, thereby providing a better candidate to estimate axonal pathways. As directional diffusion of water within the brain tends to occur along white matter tracts,  current connectome pre-processing pipelines \citep[e.g][]{cra_2013,ronc_2013} can produce an adjacency matrix $\boldsymbol{A}_i$ for each individual $i=1, \ldots, n$, with elements $A_{i[vu]}=A_{i[uv]}=1$ if there is at least one white matter fiber connecting brain regions $v=2,\ldots,V$ and $u=1, \ldots, v-1$ in individual $i$ and  $A_{i[vu]}=A_{i[uv]}=0$ otherwise. In our applications $V=68$ and each node in the network characterizes a specific anatomical brain region according to the Desikan atlas \citep{des_2006}, with the first $34$ in the left hemisphere and the remaining $34$ in the right; see Figure \ref{F1} for an illustration. Refer also to \citet{sporns_2013} for a discussion on functional and structural connectivity networks.

\subsection{Motivating application and relevant literature}
Recent studies provide brain networks along with a categorical variable. Examples include presence or absence of a neuropsychiatric disease, cognitive trait and rest-stimulus states.  There is a need for methods assessing how the brain connectivity structure varies across groups. We are specifically interested in studying the relationship between the brain connectivity structure and creative reasoning.  For each individual $i=1,\ldots,n$, data consist of an indicator of creative reasoning $y_i$ and an adjacency matrix $\boldsymbol{A}_i$ representing the undirected structural brain network.  We focus on dataset MRN-111 available at  \url{http://openconnecto.me/data/public/MR/}, preselecting subjects having high ($>111$, $y_i=2$) or low ($<90$, $y_i=1$) creative reasoning scores.  The first group comprises 19 subjects and the second 17, with thresholds chosen to correspond to the $0.15$ and $0.85$ quantiles. Creativity scores are measured via the composite creativity index (CCI) \citep{jung_2010}.  We are interested in assessing evidence of differences in brain connectivity between the low and high creativity groups, while additionally inferring the types of differences and learning which connections are responsible for these variations.  Note that we are not attempting to estimate a network, as in graphical modeling, but we are focused on testing of differences between groups in network-valued data. 

Flexible statistical methods for analyzing brain networks have lagged behind the increasingly routine collection of such data in neuroscience. A major barrier to progress in this area is that the development of statistical methodologies for formal and robust inference on network data is a challenging task. Networks represent a type of object data --- a concept encompassing a broad class of non-standard data types, ranging from functions to images and trees; refer to \citet{wang_2007} and the references cited therein for an overview. Such data require adaptations of classical modeling frameworks to non-standard spaces. This is particularly true for inference on network data in which the set of methodologies and concepts required to test for changes in underlying connectivity structures is necessarily distinct from standard data analysis strategies.

There has been some emphasis in the literature on developing methods for addressing our goals; see \citet{Bull_2009}, \citet{Stam_2014} and the references cited therein.  The main focus is on reducing each observed network $\boldsymbol{A}_i$, $i=1, \ldots, n$ to a vector of summary statistics $\boldsymbol{\theta}_i=(\theta_{i1},\ldots,\theta_{ip})^{\T}$ and then apply standard procedures, such as the multivariate analysis of variance (MANOVA), to test for changes in these vectors across groups.  Summary statistics are commonly chosen to represent global network characteristics of interest, such as the number of connections, the average path length and the clustering coefficient \citep{Rub_2010}.  Similar procedures have been recently employed in exploring the relationship between the brain network and neuropsychiatric diseases, such as Parkinson's \citep{Old_2014} and Alzheimer's \citep{Mad_2013}, but analyses are sensitive to the chosen network topological measures, with substantially different results obtained for different types of summary statistics. \citet{simps_2011} and  \citet{Simp_2012} improve choice of network summary statistics via a data driven procedure which exploits exponential random graph models \citep[e.g][]{fra_1986,wass_1996} and related validation procedures \citep{hunt_2008,hunt_20081} to detect the topological measures that better characterize the observed networks. Although this is a valuable procedure, inference is still available only on the scale of the network summary statistics, which typically discard important information about the brain connectivity architecture that may crucially explain differences across groups. Refer to \citet{ard_2010} for a review on inconsistencies in results relating brain connectivity networks to creative reasoning.  

An alternative approach is to avoid discarding information by separately testing for differences between groups in each edge probability, while adjusting the significance threshold for multiple testing via false discovery rate (FDR)  control.  As there are $V(V-1)/2$ pairs of brain regions under study --- with $V=68$ using the Desikan atlas \citep{des_2006} --- the number of tests is  substantial.  Such massively univariate approaches do not incorporate network information, leading to low power \citep{Fornito_2013}, and underestimating the variations of the brain connections across groups.  Recent proposals try to gain power by replacing the common \citet{ben_1995} approach, with thresholding procedures that account for the network structure in the data \citep{Zalesky_2010}.  However, such approaches require careful interpretation, while being highly computationally intensive, requiring permutation testing and choice of suprathreshold links. Instead of controlling FDR thresholds, \citet{scott_2014} gain power in multiple testing by using auxiliary data --- such as spatial proximity --- to inform the posterior probability that specific pairs of nodes interact differently across groups or with respect to a baseline. \citet{gine_2014} focus instead on assessing evidence of global changes in the brain structure by testing for group differences in the expected Laplacians.

 \citet{scott_2014} and  \citet{gine_2014} substantially improve state of the art in local and global hypothesis testing for network data, respectively, but are characterized by a similar key issue, motivating our methodology. Specifically, previous procedures test for changes across groups in marginal \citep{scott_2014} or expected \citep{gine_2014} structures associated with the network-valued random variable, and hence cannot detect variations in the probabilistic generative mechanism that go beyond their focus. Similarly to much simpler settings, substantially different joint probability mass functions (pmf) for a network-valued random variable can have equal expectation or induce the same marginal distributions --- characterized by the edge probabilities. Hence, these procedures are expected to fail in scenarios where the changes in the network-valued random variable are due to variations in more complex functionals.  Model misspecification can have a major effect on the quality of inference \citep{deeg_1976,begg_1990,diri_2001}, providing biased  and inaccurate conclusions.
 
\subsection{Outline of our methodology}
In order to avoid the issues discussed above, it is fundamental to define a statistical model which is sufficiently flexible to accurately approximate any probabilistic generative mechanism underlying the observed data.  \citet{dur_2014} recently proposed a flexible mixture of low-rank factorizations to characterize the distribution of a network-valued random variable. We generalize their statistical model to allow the probabilistic generative mechanism associated with the brain networks to change across groups,  without reducing data to summary measures prior to statistical analysis. 

We accomplish the above goal by factorizing the joint pmf for the random variable generating data $(y_i, \boldsymbol{A}_i)$, $i=1, \ldots,n$ as the product of the marginal pmf for the categorical predictor and the conditional pmf for the network-valued random variable given the group membership defined by the categorical predictor. By modeling the collection of group-dependent pmfs for the network-valued random variable via a flexible mixture of low-rank factorizations with group-specific mixing probabilities, we develop a simple global test for assessing evidence of group differences in the entire distribution of the network-valued random variable, rather than focusing inference only on changes in selected functionals. Differently from  \citet{gine_2014}, our procedure additionally incorporates local testing for changes in edge probabilities across groups, in line with  \citet{scott_2014} --- which in turn do not consider global tests. By explicitly borrowing strength within the network via matrix factorizations, we substantially improve power in our multiple local tests compared to standard FDR control procedures.

In Section \ref{sec:model} we describe the model formulation, with a key focus on the associated testing procedures. Prior specification, theoretical properties and posterior computation are considered in Section  \ref{sec:prior}. Section  \ref{sec:sim} provides simulations to assess inference and testing  performance of our procedures. Results for our motivating neuroscience application are discussed in Section  \ref{sec:app}. Concluding remarks are provided in Section  \ref{sec:disc}.

\section{Model formulation and testing}\label{sec:model}
\subsection{Notation and motivation}
Let $(y_i, \boldsymbol{A}_i)$ represent the creativity group and the undirected network observation, respectively, for subject $i=1, \ldots, n$, with $y_i \in \mathbb{Y}=\{1, 2\}$ and $\boldsymbol{A}_i$ the $V \times V$ adjacency matrix characterizing the edges in the network. As the brain network structure is available via undirected edges and self-relationships are not of interest, we model $(y_i, \boldsymbol{A}_i)$ by focusing on the random variable $\{\mathcal{Y}, \mathcal{L}(\boldsymbol{\mathcal{A}})\}$ generating data $\{y_i, \mathcal{L}(\boldsymbol{A}_i)\}$ with $\mathcal{L}(\boldsymbol{A}_i)=(A_{i[21]}, A_{i[31]}, \ldots, A_{i[V1]}, A_{i[32]}, \ldots,  A_{i[V2]}, \ldots, A_{i[V(V-1)]})^{\T} \in \mathbb{A}_V= \{0,1\}^{V(V-1)/2}$ the vector encoding the lower triangular elements of $\boldsymbol{A}_i$, which uniquely define the network as $A_{i[vu]}=A_{i[uv]}$ for every $v=2, \ldots, V$, $u=1, \ldots, v-1$ and $i=1, \ldots, n$.

Let ${p}_{\mathcal{Y}, \mathcal{L}(\boldsymbol{\mathcal{A}})}=\{{p}_{\mathcal{Y}, \mathcal{L}(\boldsymbol{\mathcal{A}})}(y,\boldsymbol{a}):  y \in  \mathbb{Y}, \boldsymbol{a} \in \mathbb{A}_V\}$ denote the joint pmf for the random variable $\{\mathcal{Y}, \mathcal{L}(\boldsymbol{\mathcal{A}})\}$ with ${p}_{\mathcal{Y}, \mathcal{L}(\boldsymbol{\mathcal{A}})}(y,\boldsymbol{a})=\mbox{pr}\{\mathcal{Y}=y,  \mathcal{L}(\boldsymbol{\mathcal{A}})=\boldsymbol{a}\}$, $y \in  \mathbb{Y}$ and $\boldsymbol{a} \in \mathbb{A}_V$ a network configuration.  Assessing evidence of global association between $\mathcal{Y}$ and $\mathcal{L}(\boldsymbol{\mathcal{A}})$ --- under the above notation --- formally requires testing the global null hypothesis 
\begin{eqnarray}
H_0: {p}_{\mathcal{Y}, \mathcal{L}(\boldsymbol{\mathcal{A}})}(y,\boldsymbol{a})= {p}_{\mathcal{Y}}(y){p}_{\mathcal{L}(\boldsymbol{\mathcal{A}})}(\boldsymbol{a}),
\label{global_test}
\end{eqnarray}
for all $y \in  \mathbb{Y}$ and $\boldsymbol{a} \in \mathbb{A}_V$, versus the alternative
\begin{eqnarray}
H_1: {p}_{\mathcal{Y}, \mathcal{L}(\boldsymbol{\mathcal{A}})}(y,\boldsymbol{a})\neq {p}_{\mathcal{Y}}(y){p}_{\mathcal{L}(\boldsymbol{\mathcal{A}})}(\boldsymbol{a}),
\label{global_test_1}
\end{eqnarray}
for some $y \in  \mathbb{Y}$ and $\boldsymbol{a} \in \mathbb{A}_V$, where ${p}_{\mathcal{Y}}(y)=\mbox{pr}(\mathcal{Y}=y)$, $y \in  \mathbb{Y}$ characterizes the marginal pmf of the grouping variable, whereas ${p}_{\mathcal{L}(\boldsymbol{\mathcal{A}})}(\boldsymbol{a})=\mbox{pr}\{\mathcal{L}(\boldsymbol{\mathcal{A}}) =\boldsymbol{a}\}$, $\boldsymbol{a} \in \mathbb{A}_V$ denotes the unconditional pmf for the network-valued random variable. The system of hypotheses \eqref{global_test}--\eqref{global_test_1} assesses evidence of global changes in the entire probability mass function, rather than on selected functionals or summary statistics, and hence  is more general than \citet{gine_2014} and joint tests on network measures.

Recalling our neuroscience application, rejection of $H_0$ implies that there are differences in the brain architecture across creativity groups, but fails to provide insights on the reasons for these variations. Global differences may be attributable to several underlying mechanisms, including changes in specific interconnection circuits.  As discussed in Section \ref{sec:intro}, local testing of group differences in the edge probabilities is of key interest in neuroscience applications in highlighting which brain connection measurements $\mathcal{L}(\mathcal{A})_l \in \{0,1 \}$, $l=1, \ldots, V(V-1)/2$ --- characterizing the marginals of $\mathcal{L}(\boldsymbol{\mathcal{A}})$ --- are potentially responsible for the global association between $\mathcal{Y}$ and $\mathcal{L}(\boldsymbol{\mathcal{A}})$. Hence, consistently with these interests, we also incorporate in our analyses the multiple local tests assessing --- for each pair $l=1, \ldots, V(V-1)/2$ --- evidence against the null hypothesis of independence between $\mathcal{L}(\mathcal{A})_l $ and $ \mathcal{Y}$
\begin{eqnarray}
H_{0l}: {p}_{\mathcal{Y}, \mathcal{L}({\mathcal{A}})_l}(y, a_l)={p}_{\mathcal{Y}}(y){p}_{\mathcal{L}({\mathcal{A}})_l}(a_l),
\label{local_test}
\end{eqnarray}
for all $y \in  \mathbb{Y}$ and $a_l \in \{0,1\}$, versus the alternative
\begin{eqnarray}
H_{1l}: {p}_{\mathcal{Y}, \mathcal{L}({\mathcal{A}})_l}(y, a_l)\neq {p}_{\mathcal{Y}}(y){p}_{\mathcal{L}({\mathcal{A}})_l}(a_l),
\label{local_test_1}
\end{eqnarray}
for some $y \in  \mathbb{Y}$ and $a_l \in \{0,1\}$. In hypotheses \eqref{local_test}--\eqref{local_test_1}, the quantity $p_{\mathcal{Y}, \mathcal{L}({\mathcal{A}})_l}(y,a_l)$ denotes $\mbox{pr}\{\mathcal{Y}=y, \mathcal{L}({\mathcal{A}})_l=a_l\}$, while $p_{\mathcal{L}({\mathcal{A}})_l}(a_l)=\mbox{pr}\{\mathcal{L}({\mathcal{A}})_l=a_l\}$.

In order to develop robust methodologies to test the global system \eqref{global_test}--\eqref{global_test_1}, and the multiple locals \eqref{local_test}--\eqref{local_test_1},  it is fundamental to consider a representation for ${p}_{\mathcal{Y}, \mathcal{L}(\boldsymbol{\mathcal{A}})}$ which is  provably flexible in approximating any joint pmf for data $(y_i, \boldsymbol{A}_i)$, $i=1, \ldots,n$. As $ \mathcal{L}(\boldsymbol{\mathcal{A}})$ is a highly multidimensional variable on a non-standard space, we additionally seek to reduce dimensionality in characterizing ${p}_{\mathcal{Y}, \mathcal{L}(\boldsymbol{\mathcal{A}})}$, while looking for a representation which facilitates simple derivation of  $p_{\mathcal{Y}, \mathcal{L}({\mathcal{A}})_l}(y,a_l)$ and $p_{ \mathcal{L}({\mathcal{A}})_l}(a_l)$ from ${p}_{\mathcal{Y}, \mathcal{L}(\boldsymbol{\mathcal{A}})}$. 

\subsection{Dependent mixture of low-rank factorizations}
According to the goals described above, we start by factorizing ${p}_{\mathcal{Y}, \mathcal{L}(\boldsymbol{\mathcal{A}})}$ as
\begin{eqnarray}
p_{\mathcal{Y}, \mathcal{L}(\boldsymbol{\mathcal{A}})}(y,\boldsymbol{a})= p_{\mathcal{Y}}(y)p_{\mathcal{L}(\boldsymbol{\mathcal{A}}) \mid y}(\boldsymbol{a})=\mbox{pr}(\mathcal{Y}=y)\mbox{pr}\{\mathcal{L}(\boldsymbol{\mathcal{A}}) =\boldsymbol{a} \mid \mathcal{Y}=y\},
\label{eq1}
\end{eqnarray}
for every $y \in \mathbb{Y}$ and $\boldsymbol{a} \in \mathbb{A}_V$. It is always possible to define the joint probability mass function ${p}_{\mathcal{Y}, \mathcal{L}(\boldsymbol{\mathcal{A}})}$ as the product of the marginal pmf ${p}_{\mathcal{Y}}=\{{p}_{\mathcal{Y}}(y): y \in \mathbb{Y} \}$ for the grouping variable and the conditional pmfs ${p}_{\mathcal{L}(\boldsymbol{\mathcal{A}}) \mid y}=\{p_{\mathcal{L}(\boldsymbol{\mathcal{A}}) \mid y}(\boldsymbol{a}): \boldsymbol{a} \in \mathbb{A}_V \}$ for the network-valued random variable given the group $y \in \mathbb{Y}$. This also favors inference on how the network structure varies across the two groups, with ${p}_{\mathcal{L}(\boldsymbol{\mathcal{A}}) \mid 1}$ and ${p}_{\mathcal{L}(\boldsymbol{\mathcal{A}}) \mid 2}$ fully characterizing such variations. Although we treat $\mathcal{Y}$ as a random variable through a prospective likelihood, our methodology remains also valid for  studies that sample the groups under a retrospective design. 

Under factorization \eqref{eq1}, the global test \eqref{global_test}--\eqref{global_test_1} coincides with assessing whether the conditional pmf of the network-valued random variable remains equal or shifts across the two groups. Hence, under  \eqref{eq1} hypotheses \eqref{global_test}--\eqref{global_test_1} reduce to
\begin{eqnarray}
 H_0: {p}_{\mathcal{L}(\boldsymbol{\mathcal{A}}) \mid 1}(\boldsymbol{a})= {p}_{\mathcal{L}(\boldsymbol{\mathcal{A}}) \mid 2}(\boldsymbol{a}), \quad \mbox{for all} \ \boldsymbol{a} \in \mathbb{A}_V,
\label{global_test1} 
\end{eqnarray}
versus the alternative
\begin{eqnarray}
\ \ \ \ H_1: {p}_{\mathcal{L}(\boldsymbol{\mathcal{A}}) \mid 1}(\boldsymbol{a})\neq {p}_{\mathcal{L}(\boldsymbol{\mathcal{A}}) \mid 2}(\boldsymbol{a}), \quad \mbox{for some} \ \boldsymbol{a} \in \mathbb{A}_V.
\label{global_test2} 
\end{eqnarray}
In order to develop a provably general and robust strategy to test \eqref{global_test1}--\eqref{global_test2} the key challenge relies in flexibly modeling the conditional pmfs ${p}_{\mathcal{L}(\boldsymbol{\mathcal{A}}) \mid 1}$ and ${p}_{\mathcal{L}(\boldsymbol{\mathcal{A}}) \mid 2}$ characterizing the distribution of the network-valued random variable  in the first and second group, respectively. In fact, for every group $y \in \mathbb{Y}$, one needs a parameter $p_{\mathcal{L}(\boldsymbol{\mathcal{A}}) \mid y}(\boldsymbol{a})$ for every possible network configuration $\boldsymbol{a} \in \mathbb{A}_V$ to uniquely characterize ${p}_{\mathcal{L}(\boldsymbol{\mathcal{A}}) \mid y}$, with the number of configurations being $|\mathbb{A}_V|=2^{V(V-1)/2}$. For example, in our neuroscience application  $|\mathbb{A}_{68}|=2^{68(68-1)/2}-1=2^{2{,}278}-1$ free parameters are required to uniquely define the pmf of the brain networks in each group $y \in \mathbb{Y}$ under the usual restriction $\sum_{\boldsymbol{a} \in \mathbb{A}_{68}} p_{\mathcal{L}(\boldsymbol{\mathcal{A}}) \mid y}(\boldsymbol{a})=1$. Clearly this number of parameters to test is massively larger than the sample size available in neuroscience applications. Hence, to facilitate tractable testing procedures it is necessary to substantially reduce dimensionality.  However, in reducing dimension, it is
important to avoid making overly restrictive assumptions that  lead to formulations sensitive to issues arising from model misspecification.

Focused on modeling a network-valued random variables' pmf, ${p}_{\mathcal{L}(\boldsymbol{\mathcal{A}})}$, without considering hypothesis testing or additional data on a categorical response,  \citet{dur_2014} proposed a mixture of low-rank factorizations which reduces dimensionality by exploiting network information while retaining flexibility.  Although this provides an appealing building block for our testing procedures, global and local testing and inference on group differences is not a straightforward add on to their approach.  As a first step towards constructing our tests, we generalize their model to allow group differences via
\begin{eqnarray}
p_{\mathcal{L}(\boldsymbol{\mathcal{A}}) \mid y}(\boldsymbol{a})=\mbox{pr}\{\mathcal{L}(\boldsymbol{\mathcal{A}}) =\boldsymbol{a} \mid \mathcal{Y}=y\}=  \sum_{h=1}^{H} \nu_{hy} \prod_{l=1}^{V(V-1)/2} (\pi_{l}^{(h)})^{a_l} (1-  \pi^{(h)}_{l})^{1-a_l}, 
\label{eq2}
\end{eqnarray}
for each configuration $\boldsymbol{a} \in \mathbb{A}_{V}$ and group $y \in \{1,2\}$,  with the edge probability vectors $\boldsymbol{\pi}^{(h)}=(\pi_1^{(h)}, \ldots, \pi_{V(V-1)/2}^{(h)} )^{\T} \in (0,1)^{V(V-1)/2}$ in each mixture component, defined as
\begin{eqnarray}
\boldsymbol{\pi}^{(h)}=\left\{1+\exp({-\boldsymbol{Z}-\boldsymbol{D}^{(h)}})\right\}^{-1},\quad \boldsymbol{D}^{(h)}=\mathcal{L}(\boldsymbol{X}^{(h)}\boldsymbol{\Lambda}^{(h)} \boldsymbol{X}^{(h)\T}), \quad h=1, \ldots, H, 
\label{eq2_1}
\end{eqnarray}
with $\boldsymbol{X}^{(h)} \in \Re^{V\times R}$, $\boldsymbol{\Lambda}^{(h)}$ diagonal with $R$ non-negative weights $\lambda^{(h)}_{1}, \ldots, \lambda^{(h)}_{R}$, and  $\boldsymbol{Z} \in \Re^{V(V-1)/2}$. Representation \eqref{eq2} defines ${p}_{\mathcal{L}(\boldsymbol{\mathcal{A}}) \mid 1}$ and ${p}_{\mathcal{L}(\boldsymbol{\mathcal{A}}) \mid 2}$ via a flexible dependent mixture model, which borrows strength across the two groups in characterizing the shared mixture components, while allowing flexible modeling of the conditional pmfs $p_{\mathcal{L}(\boldsymbol{\mathcal{A}}) \mid y}$ via group-specific mixing probabilities $\boldsymbol{\nu}_{y}=(\nu_{1y}, \ldots, \nu_{Hy})$,  $y \in \{1,2\}$, with $\nu_{hy} \in (0,1)$ for all $h=1, \ldots, H$ and $\sum_{h=1}^H\nu_{hy} =1$ for every $y \in \{1,2\}$. 

In order to reduce dimensionality and efficiently borrow information within the network, the characterization of the mixture components in \eqref{eq2_1} adapts concepts from the literature on latent variable modeling of networks.  Refer to  \citet{now_2001}, \citet{air_2008}, \citet{hof_2002} and \citet{hof_2008} for popular specifications in modeling of a single network observation. Within each mixture component, connections among pairs of nodes are characterized as conditionally independent Bernoulli random variables given their component-specific edge probabilities $\pi_l^{(h)}$, $l=1, \ldots, V(V-1)/2$, with these probabilities further characterized as a function of node-specific latent variables. In equation \eqref{eq2_1}, we define each component-specific log-odds vector as the sum of a shared similarity  $\boldsymbol{Z} \in \Re^{V(V-1)/2}$ and a component-specific one $\boldsymbol{D}^{(h)} \in \Re^{V(V-1)/2}$ arising from the weighted dot product of node-specific latent coordinate vectors defining the rows of the $V \times R$ --- typically $R \ll V$ --- matrix $\boldsymbol{X}^{(h)}$, for $h=1, \ldots, H$. In fact, letting $l$ denote the pair of nodes $v$ and $u$, $v>u$, under  \eqref{eq2_1}, the probability of an edge between $v$ and $u$ in component $h$ increases with $Z_l$ and $\sum_{r=1}^{R}\lambda^{(h)}_r X^{(h)}_{vr} X^{(h)}_{ur}$. Representation  \eqref{eq2_1} provides an over-complete factorization --- a common approach providing several benefits in Bayesian hierarchical modeling of multidimensional data \citep[e.g][]{bha_2011,ghosh_2009}. Factorization \eqref{eq2_1} is appealing in reducing dimensionality, accommodating key topological network properties \citep{hof_2008} and improving mixing performance \citep{gelm_2012}. Our focus is on  using the resulting flexible and tractable formulation \eqref{eq2}--\eqref{eq2_1} to draw inference on changes in identified functionals of interest arising from the pmf of our network-valued random variable and develop robust procedures for global and local testing.

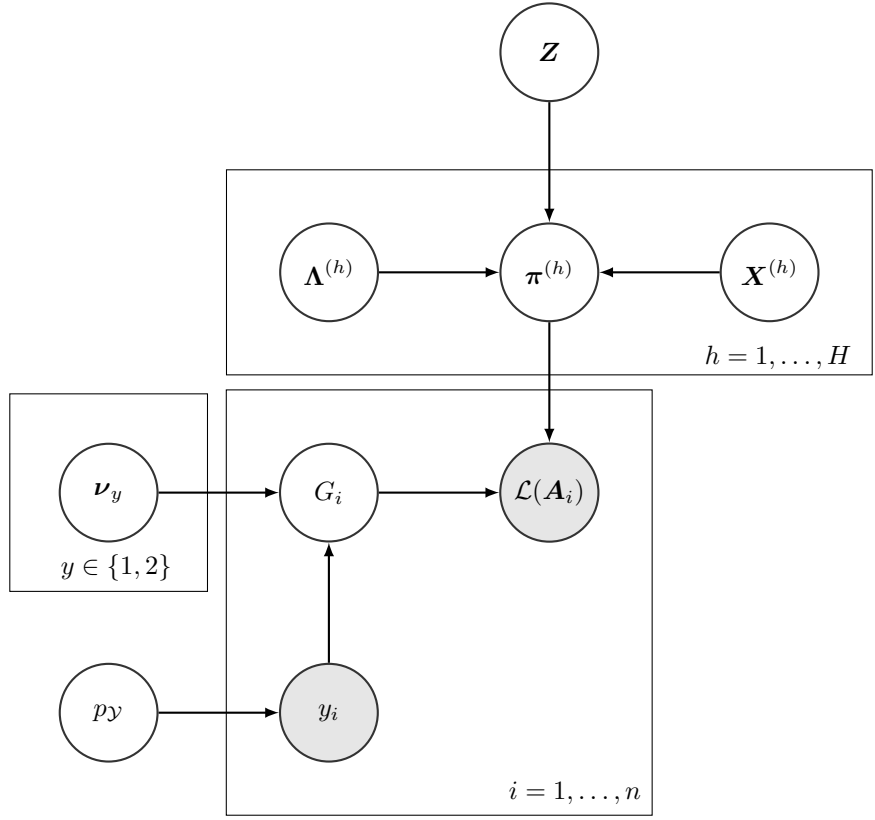
\begin{figure}[t]
\centering
\begin{tikzpicture}
\tikzstyle{main}=[circle, minimum size = 13mm, thick, draw =black!80, node distance = 16mm]
\tikzstyle{connect}=[-latex, thick]
\tikzstyle{box}=[rectangle, draw=black!100]
  \node[main, fill = white!100] (theta) {$\boldsymbol{\nu}_y$ };
  \node[main] (z) [right=of theta] {$G_i$};
    \node[main,fill = black!10] (d) [below=of z] {$y_i$ };
        \node[main] (p) [left=of d] {${p}_{\mathcal{Y}}$ };
  \node[main, fill = black!10] (w) [right=of z] {$\mathcal{L}(\boldsymbol{A}_i)$ };
    \node[main] (pi) [above=of w] {$\boldsymbol{\pi}^{(h)}$ };
        \node[main] (x) [right=of pi] {$\boldsymbol{X}^{(h)}$ };
                \node[main] (la) [left=of pi] {$\boldsymbol{\Lambda}^{(h)}$ };
                                \node[main] (zeta) [above=of pi] {$\boldsymbol{Z}$ };
     \path        (theta) edge [connect] (z)
        (z) edge [connect] (w)
        (pi) edge [connect] (w)
        (d) edge [connect] (z)
                (p) edge [connect] (d)
                  (x) edge [connect] (pi)
                    (la) edge [connect] (pi)
                      (zeta) edge [connect] (pi);
  \node[rectangle, inner sep=6.5mm, draw=black!100, fit = (theta)] {};
  \node[rectangle, inner sep=0mm, fit= (theta)] {};
  \node[rectangle, inner sep=7mm, draw=black!100, fit = (d) (z) (w)] {};
    \node[rectangle, inner sep=7mm, draw=black!100, fit = (pi) (la) (x)] {};
  \node[] at (0.1,-1) {$y \in \{1, 2\}$};
    \node[] at (6.2,-4) {$i=1, \ldots, n$};
        \node[] at (8.9,1.8) {$h=1, \ldots, H$};
\end{tikzpicture}
\caption{Graphical representation of the mechanism to generate data $\{y_i, \mathcal{L}(\boldsymbol{A}_i)\}$, $i=1, \ldots,n$ under representation \eqref{eq1} and \eqref{eq2}--\eqref{eq2_1} for the joint pmf ${p}_{\mathcal{Y}, \mathcal{L}(\boldsymbol{\mathcal{A}})}$.}\label{F_dep}
\end{figure}

Figure \ref{F_dep} outlines the mechanism to generate data $\{y_i, \mathcal{L}(\boldsymbol{A}_i)\}$ from the random variable $\{\mathcal{Y}, \mathcal{L}(\boldsymbol{\mathcal{A}})\}$ with pmf factorized as in  \eqref{eq1} and \eqref{eq2}--\eqref{eq2_1}. According to Figure \ref{F_dep} the indicator group $y_i$ is sampled from ${p}_{\mathcal{Y}}$. The network $\mathcal{L}(\boldsymbol{A}_i)$ is instead generated conditioned on $y_i$ under the mixture representation in \eqref{eq2}. In particular, given $y_i=y$ we first choose a mixture component by sampling the latent indicator $G_i \in \{1, \ldots, H\}$ with conditional pmf defined by the mixing probabilities, so that ${p}_{G_i \mid y}(h)=\nu_{hy}$. Then, given $G_i=h$ and the corresponding edge probability vector $\boldsymbol{\pi}^{(h)}$ --- factorized as in  \eqref{eq2_1} --- the network $ \mathcal{L}(\boldsymbol{A}_i)$ is generated by sampling its edges $\mathcal{L}({A}_i)_l$, $l=1,\ldots, V(V-1)/2$ from conditionally independent Bernoulli variables. Hence, the dependence on the groups is introduced in the assignments to the mixture components via group-specific mixing probabilities, so that brain networks in the same component share a common edge probability vector $\boldsymbol{\pi}^{(h)}$, with the probability assigned to each component changing across the two groups. This simple generative mechanism is appealing in facilitating tractable posterior computation and inference.

A key aspect in representation   \eqref{eq2}--\eqref{eq2_1} is that it allows dimensionality reduction, while preserving flexibility. As stated in Proposition \ref{lem1} such a representation is sufficiently flexible to characterize any collection of group-dependent pmfs ${p}_{\mathcal{L}(\boldsymbol{\mathcal{A}}) \mid 1}$, ${p}_{\mathcal{L}(\boldsymbol{\mathcal{A}}) \mid 2}$.
\begin{prop}
Any collection of group-dependent probability mass functions ${p}_{\mathcal{L}(\boldsymbol{\mathcal{A}}) \mid y} \in \mathcal{P}_{| \mathbb{A}_V |}=\{{p}_{\mathcal{L}(\boldsymbol{\mathcal{A}}) \mid y}: 0\leq{p}_{\mathcal{L}(\boldsymbol{\mathcal{A}}) \mid y}(\boldsymbol{a})\leq 1 \ \mbox{for all} \ \boldsymbol{a} \in \mathbb{A}_V,  \sum_{\boldsymbol{a} \in \mathbb{A}_V}{p}_{\mathcal{L}(\boldsymbol{\mathcal{A}}) \mid y}(\boldsymbol{a})=1\}$, $y \in \{1,2\}$ can be characterized as in \eqref{eq2} for some $H$ with component-specific edge probability vectors $\boldsymbol{\pi}^{(h)}$, $h=1, \ldots, H$ factorized as in \eqref{eq2_1} for some $R$.
\label{lem1}
\end{prop}
This additionally ensures that any joint probability mass function ${p}_{\mathcal{Y}, \mathcal{L}(\boldsymbol{\mathcal{A}})}$  for the random variable $\{\mathcal{Y}, \mathcal{L}(\boldsymbol{\mathcal{A}})\}$ admits representation \eqref{eq1}, \eqref{eq2}--\eqref{eq2_1} and hence our formulation can be viewed as fully general and robust against model misspecification in testing \eqref{global_test1}--\eqref{global_test2}, given sufficiently flexible priors for the components. See the online supplementary materials for proofs.

\subsection{Global and local testing under the proposed statistical model}
Including group dependence only in the mixing probabilities favors borrowing of information across the groups in modeling $\boldsymbol{\pi}^{(h)}$, $h=1, \ldots, H$, while massively reducing the number of parameters to test in \eqref{global_test1}--\eqref{global_test2} from  $2\{2^{V(V-1)/2}-1\}$  to $2(H-1)$. In fact, the characterization of ${p}_{\mathcal{L}(\boldsymbol{\mathcal{A}}) \mid y}$ in \eqref{eq2}--\eqref{eq2_1} further simplifies the system \eqref{global_test1}--\eqref{global_test2} to only testing the equality of the group-specific mixing probability vectors
\begin{eqnarray}
H_0: (\nu_{11},\ldots, \nu_{H1})=(\nu_{12},\ldots, \nu_{H2}) \ \  \mbox{versus} \ \ H_1: (\nu_{11},\ldots, \nu_{H1})\neq (\nu_{12},\ldots, \nu_{H2}). \ \
\label{eq_mix_test}
\end{eqnarray}
Recalling Proposition \ref{lem1}, under our  formulation, the system \eqref{eq_mix_test} uniquely characterizes the global hypotheses  \eqref{global_test}--\eqref{global_test_1}.

In developing methodologies for the multiple local tests in \eqref{local_test}--\eqref{local_test_1} under our model formulation, we measure the association between $\mathcal{L}(\mathcal{A})_l$ and $ \mathcal{Y}$ exploiting the model-based version of the Cramer's V proposed in \citet{dun_2009}, obtaining
\begin{eqnarray}
\rho^2_{l}&=&\frac{1}{\mbox{min}\{2,2\}-1}\sum_{y=1}^2 \sum_{a_l=0}^1\frac{\left\{p_{\mathcal{Y}, \mathcal{L}({\mathcal{A}})_l}(y,a_l)- p_{\mathcal{Y}}(y)p_{\mathcal{L}({\mathcal{A}})_l}(a_l)\right\}^2}{p_{\mathcal{Y}}(y)p_{\mathcal{L}({\mathcal{A}})_l}(a_l)}\nonumber \\
&=&\sum_{y=1}^2 \sum_{a_l=0}^1\frac{\left\{p_{\mathcal{Y}}(y)p_{\mathcal{L}({\mathcal{A}})_l \mid y}(a_l)- p_{\mathcal{Y}}(y)p_{\mathcal{L}({\mathcal{A}})_l}(a_l)\right\}^2}{p_{\mathcal{Y}}(y)p_{\mathcal{L}({\mathcal{A}})_l}(a_l)}\nonumber \\
&=&\sum_{y=1}^2 p_{\mathcal{Y}}(y) \sum_{a_l=0}^1\frac{\left\{p_{\mathcal{L}({\mathcal{A}})_l \mid y}(a_l)-p_{\mathcal{L}({\mathcal{A}})_l }(a_l)\right\}^2}{p_{\mathcal{L}({\mathcal{A}})_l }(a_l)}.
\label{ro}
\end{eqnarray}
Measuring the local association with $\rho_{l} \in (0,1)$ provides an appealing choice in terms of interpretation, with $\rho_{l} = 0$ meaning that ${p}_{\mathcal{Y}, \mathcal{L}({\mathcal{A}})_l}(y, a_l)={p}_{\mathcal{Y}}(y){p}_{\mathcal{L}({\mathcal{A}})_l}(a_l)$, for all $y \in  \mathbb{Y}$ and $a_l \in \{0,1\}$, and hence the random variable $\mathcal{L}(\mathcal{A})_l$ modeling the presence or absence of an edge among the $l$th pair of nodes, has no differences across groups. Beside incorporating a fully general and tractable global test, our model formulation is particularly appealing also in addressing issues associated to local multiple testing in the network framework. First, as stated in Proposition \ref{lem2.1}, each $\rho_l$, $l=1, \ldots, V(V-1)/2$, can be easily computed from the quantities in our model. 
\begin{prop}
Based on factorizations \eqref{eq1} and \eqref{eq2}, $p_{\mathcal{L}({\mathcal{A}})_l \mid y}(1)=1-p_{\mathcal{L}({\mathcal{A}})_l \mid y}(0)=\sum_{h=1}^{H}\nu_{hy}\pi^{(h)}_l$, and $p_{\mathcal{L}({\mathcal{A}})_l }(1)=1-p_{\mathcal{L}({\mathcal{A}})_l }(0)=\sum_{y=1}^{2}p_{\mathcal{Y}}(y)\sum_{h=1}^{H}\nu_{hy}\pi^{(h)}_l$.
\label{lem2.1}
\end{prop}
Second, the shared dependence on a common set of node-specific latent coordinates characterizing the construction of the edge probability vector $\boldsymbol{\pi}^{(h)}$ within each mixture component $h=1, \ldots, H$ in  \eqref{eq2_1}, explicitly accounts for specific dependence structures in brain connections. According to \citet{hof_2008}, factorization  \eqref{eq2_1} can accurately accommodate key topological properties including block structures, homophily behaviors and transitive edge patterns --- among others. As a result --- in line with \citet{scott_2014} --- informing our local testing procedures about these structures, is expected to substantially improve power compared to standard FDR control procedures.

\section{Prior specification and posterior computation}
\label{sec:prior}
\subsection{Prior specification and properties}
We specify independent priors ${p}_{\mathcal{Y}}\sim\Pi_{y}$, $\boldsymbol{Z}=(Z_1, \ldots, Z_{V(V-1)/2})^{\T} \sim \Pi_Z$, $\boldsymbol{X}^{(h)} \sim \Pi_{X}$, $\boldsymbol{\lambda}^{(h)}=(\lambda^{(h)}_{1}, \ldots, \lambda^{(h)}_{R})^{\T} \sim \Pi_{\lambda}$,  $h=1, \ldots, H$ and $\boldsymbol{\nu}_{y}=(\nu_{1y}, \ldots, \nu_{Hy})\sim \Pi_{\nu}$, $y \in \{1, 2\}$, to induce a prior $\Pi$ on the joint pmf ${p}_{\mathcal{Y}, \mathcal{L}(\boldsymbol{\mathcal{A}})}$ with full support in $\mathcal{P}_{2 \times | \mathbb{A}_V |}=\{{p}_{\mathcal{Y},\mathcal{L}(\boldsymbol{\mathcal{A}})}: 0\leq{p}_{\mathcal{Y},\mathcal{L}(\boldsymbol{\mathcal{A}})}(y,\boldsymbol{a})\leq 1 \ \mbox{for all} \ y \in \{1,2\}, \boldsymbol{a} \in \mathbb{A}_V,  \ \mbox{with} \ \sum_{y \in \{1,2\}, \boldsymbol{a} \in \mathbb{A}_V}{p}_{\mathcal{Y},\mathcal{L}(\boldsymbol{\mathcal{A}})}(y,\boldsymbol{a})=1\}$, while obtaining desirable asymptotic behavior, simple posterior computation and allowance for testing. Prior support is a key property to retain the flexibility associated with our statistical model and testing procedures, when performing posterior inference.

As ${p}_{\mathcal{Y}}$ is the pmf for a categorical variable on $2$ levels, we let $1-p_{\mathcal{Y}}(2)=p_{\mathcal{Y}}(1) \sim \mbox{Beta}(a, b)$, and consider the same prior specification suggested by \citet{dur_2014} for the quantities in \eqref{eq2_1} by choosing Gaussian priors for the entries in $\boldsymbol{Z}$, standard Gaussians for the elements in the coordinates matrix $\boldsymbol{X}^{(h)}$ and multiplicative inverse gammas for $\boldsymbol{\lambda}^{(h)} \sim \mbox{MIG}(a_1,a_2)$, $h=1, \ldots, H$, \citep{bha_2011}. This choice for $\Pi_{\lambda}$ favors shrinkage, with elements in $\boldsymbol{\lambda}^{(h)}$ increasingly concentrated close to $0$ as $r$ increases, so as to shrink towards lower dimensional representations and adaptively penalize high dimensional ones. A key property of our prior specification is incorporation of global testing \eqref{eq_mix_test} in the definition of $\Pi_{\nu}$. Specifically letting $\boldsymbol{\upsilon}=(\upsilon_1, \ldots, \upsilon_H)$ and $\boldsymbol{\upsilon}_y=(\upsilon_{1y}, \ldots, \upsilon_{Hy})$, we induce $\Pi_{\nu}$ through
\begin{eqnarray}
\boldsymbol{\nu}_{y} &=&(1-T)\boldsymbol{\upsilon}+T\boldsymbol{\upsilon}_{y}, \quad y\in \{1, 2\},\nonumber\\
\boldsymbol{\upsilon} &\sim& \mbox{Dir}(1/H, \ldots, 1/H), \quad \boldsymbol{\upsilon}_{y} \sim  \mbox{Dir}(1/H, \ldots, 1/H), \ y\in \{1, 2\}, \label{eq5} \\ 
T &\sim& \mbox{Bern}\{\mbox{pr}(H_1)\}.\nonumber
\end{eqnarray}
In (\ref{eq5}), $T$ is a hypothesis indicator, with $T=0$ for $H_0$ and $T=1$ for $H_1$.  Under $H_1$, we generate group-specific mixing probabilities independently, while under $H_0$ we have equal probability vectors.  By choosing small values for the parameters in the Dirichlet priors, we favor automatic deletion of redundant components \citep{rou_2011}.  In assessing evidence in favor of the alternative, we can rely on the posterior probability, $\mbox{pr}[H_1 \mid \{\boldsymbol{y},\mathcal{L}(\boldsymbol{A})\}]=1-\mbox{pr}[H_0 \mid \{\boldsymbol{y},\mathcal{L}(\boldsymbol{A})\}]$ which can be easily obtained from the output of the Gibbs sampler proposed below. Specifically, under prior \eqref{eq5} and exploiting the hierarchical structure of our dependent mixture model --- summarized in Figure \ref{F_dep} --- the full conditional $\mbox{pr}(T=1 \mid -)=\mbox{pr}(H_1 \mid -)=1-\mbox{pr}(H_0 \mid -)$ is 
{\small{\begin{eqnarray}
\mbox{pr}(H_1 \mid -)&=& \frac{\mbox{pr}(H_1)\prod_{y=1}^{2}\int (\prod_{h=1}^{H}\upsilon_{hy}^{n_{hy}})d \Pi_{\upsilon_y} }{\mbox{pr}(H_0)\int (\prod_{h=1}^{H}\upsilon_h^{n_h})d \Pi_\upsilon +\mbox{pr}(H_1)\prod_{y=1}^{2}\int (\prod_{h=1}^{H}\upsilon_{hy}^{n_{hy}})d \Pi_{\upsilon_y} },\nonumber\\
&=&\frac{\mbox{pr}(H_1)\prod_{y=1}^2 \{\mbox{B}(\boldsymbol{\alpha}+\boldsymbol{\bar{n}}_y)/  \mbox{B}(\boldsymbol{\alpha})\}}{\mbox{pr}(H_0) \mbox{B}(\boldsymbol{\alpha}+\boldsymbol{\bar{n}})/  \mbox{B}(\boldsymbol{\alpha})+\mbox{pr}(H_1)\prod_{y=1}^2 \{ \mbox{B}(\boldsymbol{\alpha}+\boldsymbol{\bar{n}}_y)/  \mbox{B}(\boldsymbol{\alpha})\}},
\label{eq6}
\end{eqnarray}}}with $n_{hy}=\sum_{i:y_i=y}\mbox{1}(G_i=h)$, $n_h=\sum_{i=1}^{n}\mbox{1}(G_i=h)$,  $\boldsymbol{\bar{n}}_y=(n_{1y}, \ldots, n_{Hy})$, $\boldsymbol{\bar{n}}=(n_1, \ldots, n_H)$, $\boldsymbol{\alpha}=(1/H, \ldots, 1/H)$, and $\mbox{B}(\cdot)$ is the multivariate beta function. It is easy to derive the equalities $\int(\prod_{h=1}^{H}\upsilon_h^{n_h})d \Pi_\upsilon=\mbox{B}(\boldsymbol{\alpha}+\boldsymbol{\bar{n}})/  \mbox{B}(\boldsymbol{\alpha})$ and $\int(\prod_{h=1}^{H}\upsilon_{hy}^{n_{hy}})d \Pi_{\upsilon_y}= \mbox{B}(\boldsymbol{\alpha}+\boldsymbol{\bar{n}}_y)/  \mbox{B}(\boldsymbol{\alpha})$, $y \in \{1,2\}$ exploiting the Dirichlet-multinomial conjugacy.

Although providing a key choice for performing global testing, it is impractical to adopt formulation (\ref{eq5}) for each local point null $H_{0l}: \rho_l =0$ versus $H_{1l}: \rho_l \neq 0$, $l=1, \ldots, V(V-1)/2$. Hence, we replace local point nulls with small interval nulls $H_{0l}: \rho_l \leq \epsilon$ versus $H_{1l}: \rho_l > \epsilon$. This choice allows $\mbox{pr}[H_{1l} \mid \{\boldsymbol{y},\mathcal{L}\boldsymbol{(A)}\}]=1-\mbox{pr}[H_{0l} \mid \{\boldsymbol{y},\mathcal{L}\boldsymbol{(A)}\}]$ to be easily estimated as the proportion of Gibbs samples in which $\rho_l >\epsilon$, for each $l=1, \ldots, V(V-1)/2$. Moreover --- as noted in \citet{berg_1987} and \citet{berg_1988} --- testing the small interval hypothesis $H_{0l}: \rho_l \leq \epsilon$  is in general more realistic and provides --- under a Bayesian paradigm --- comparable results to those obtained when assessing evidence of $H_{0l}: \rho_l =0$.

Beside providing key computational benefits, as stated in Proposition \ref{lem2}, our choices induce a prior $\Pi$ for ${p}_{\mathcal{Y}, \mathcal{L}(\boldsymbol{\mathcal{A}})}$ with full $L_1$ support over  $\mathcal{P}_{2 \times | \mathbb{A}_V |}$, meaning that $\Pi$ can generate a ${p}_{\mathcal{Y}, \mathcal{L}(\boldsymbol{\mathcal{A}})}$ within an arbitrarily small $L_1$ neighborhood of the true data-generating model ${p}^0_{\mathcal{Y}, \mathcal{L}(\boldsymbol{\mathcal{A}})}$, allowing the truth to fall in a wide class.

\begin{prop}
Based on our priors $\Pi_{y}, \Pi_Z, \Pi_{X}, \Pi_{\lambda}$, $\Pi_{\nu}$, and letting $\mathbb{B}_{\epsilon}({p}^0_{\mathcal{Y}, \mathcal{L}(\boldsymbol{\mathcal{A}})})=\{{p}_{\mathcal{Y}, \mathcal{L}(\boldsymbol{\mathcal{A}})}:\sum_{y=1}^{2}  \sum_{\boldsymbol{a} \in \mathbb{A}_V} |p_{\mathcal{Y}, \mathcal{L}(\boldsymbol{\mathcal{A}})}(y,\boldsymbol{a})- p^0_{\mathcal{Y}, \mathcal{L}(\boldsymbol{\mathcal{A}})}(y,\boldsymbol{a})|< \epsilon \}$ denote the $L_1$ neighborhood around ${p}^0_{\mathcal{Y}, \mathcal{L}(\boldsymbol{\mathcal{A}})}$, then for any ${p}^0_{\mathcal{Y}, \mathcal{L}(\boldsymbol{\mathcal{A}})}\in \mathcal{P}_{2 \times | \mathbb{A}_V |}$ and $\epsilon>0$, $\Pi\{\mathbb{B}_{\epsilon}({p}^0_{\mathcal{Y}, \mathcal{L}(\boldsymbol{\mathcal{A}})})\}>0$.
\label{lem2}
\end{prop}

Full prior support is a key property to ensure accurate posterior inference and testing, because without prior support about the true data-generating pmf, the posterior cannot possibly concentrate around the truth. Moreover, as ${p}_{\mathcal{Y}, \mathcal{L}(\boldsymbol{\mathcal{A}})}$ is characterized by finitely many parameters $p_{\mathcal{Y}, \mathcal{L}(\boldsymbol{\mathcal{A}})}(y,\boldsymbol{a})$, $y\in  \mathbb{Y}$, $\boldsymbol{a} \in \mathbb{A}_V$, Proposition \ref{lem2} is sufficient to guarantee that the posterior assigns probability one to any arbitrarily small neighborhood of the true joint pmf as  $n \rightarrow \infty$, meaning that $\Pi[\mathbb{B}_{\epsilon}({p}^0_{\mathcal{Y}, \mathcal{L}(\boldsymbol{\mathcal{A}})}) \mid \{y_1,\mathcal{L}(\boldsymbol{A}_1)\}, \ldots,  \{y_n,\mathcal{L}(\boldsymbol{A}_n)\}]$ converges almost surely to $1$, when the true joint pmf is ${p}^0_{\mathcal{Y}, \mathcal{L}(\boldsymbol{\mathcal{A}})}$.

\subsection{Posterior computation}
Posterior computation is available via a simple Gibbs sampler, exploiting our representation in Figure \ref{F_dep}. Specifically, the MCMC alternates between the following steps.
\begin{enumerate}
\item Sample $p_{\mathcal{Y}}(1)=1-p_{\mathcal{Y}}(2)$ from the full conditional  $p_{\mathcal{Y}}(1) \mid -\sim \mbox{Beta}(a+n_1, b+n_2) $, with $n_y= \sum_{i=1}^n \mbox{1}(y_i=y)$.
\item
For each $i=1, \ldots, n$, update $G_i$ from the discrete variable with probabilities,
\begin{eqnarray*}
\mbox{pr}(G_i=h \mid - )= \frac{ \nu_{h y_{ _i}} \prod_{l=1}^{V(V-1)/2}(\pi_{l}^{(h)})^{ \mathcal{L}(A_i)_l} (1-  \pi^{(h)}_{l})^{1- \mathcal{L}(A_i)_l}}{\sum_{q=1}^{H}\nu_{q y_{ _i}} \prod_{l=1}^{V(V-1)/2}(\pi_{l}^{(q)})^{ \mathcal{L}(A_i)_l} (1-  \pi^{(q)}_{l})^{1- \mathcal{L}(A_i)_l}},
\end{eqnarray*}
for $h=1, \ldots, H$, with each $\boldsymbol{\pi}^{(h)}$ factorized as in \eqref{eq2_1}
\item
Given $G_i$, $i=1, \ldots, n$, the updating for quantities $\boldsymbol{Z}$, $\boldsymbol{X}^{(h)}$ and $\boldsymbol{\lambda}^{(h)}$, $h=1, \ldots, H$ proceeds via the recently developed Poly\'a-gamma data augmentation scheme for Bayesian logistic regression \citep{pol_2013} as in \citet{dur_2014}.
\item
Sample the testing indicator $T$ from a Bernoulli with probability (\ref{eq6}).
\item
If $T=0$, let $\boldsymbol{\nu}_{y}=\boldsymbol{\upsilon}$, $y\in \{1,2\}$ with $\boldsymbol{\upsilon}$ updated from the full conditional Dirichlet $(\upsilon_1, \ldots, \upsilon_H) \mid -\sim \mbox{Dir}(1/H+n_1, \ldots, 1/H+n_H)$. Otherwise, if $T=1$, update each $\boldsymbol{\nu}_{y}$ independently from $(\nu_{1y}, \ldots, \nu_{Hy}) \mid -\sim \mbox{Dir}(1/H+n_{1y}, \ldots, 1/H+n_{Hy})$. 
\end{enumerate}

Since the number of mixture components in \eqref{eq2} and the dimensions of the latent spaces in \eqref{eq2_1} are not known in practice, we perform posterior computation by fixing $H$ and $R$ at conservative upper bounds.  The priors $\Pi_{\nu}$ and $\Pi_{\lambda}$ are chosen to allow adaptive emptying of the redundant components, with the posteriors for the corresponding parameters controlling unnecessary dimensions concentrated near zero.  

\section{Simulation studies}
\label{sec:sim}
We consider simulation studies to evaluate the performance of our method in  correctly assessing the global hypothesis of association among the network-valued random variable $\mathcal{L}(\boldsymbol{\mathcal{A}})$ and the categorical predictor $\mathcal{Y}$, and in identifying local variations in each edge probability across groups.

For comparison we also implement a MANOVA  procedure  --- see e.g. \citet{krz_1988} --- to test for global variations across groups in the random vector of summary measures $\boldsymbol{\Theta}$, with realization $\boldsymbol{\theta}_i$ from $\boldsymbol{\Theta}$ comprising the most common network summary statistics --- covering network density, transitivity, average path length and assortativity --- computed for each simulated network $\boldsymbol{A}_i$. Refer to \citet{kan_2013} for an overview on these topological network measures and \citet{Bull_2009}, \citet{bull_2012} for a discussion on their importance in characterizing wiring mechanisms within the brain. For local testing, we compare our procedure to the results obtained when testing on the association between $\mathcal{L}({\mathcal{A}})_l$ and $\mathcal{Y}$  via separate two-sided Fisher's exact tests for each $l=1, \ldots, V(V-1)/2$ --- see e.g. \citet{agr_2002}. We consider exact tests to avoid issues arising from $\chi^2$ approximations in sparse tables.

\begin{figure}[t]
\centering
\includegraphics[trim=0.75cm 0cm 0.22cm 0cm, clip=true,height=7cm, width=13.2cm]{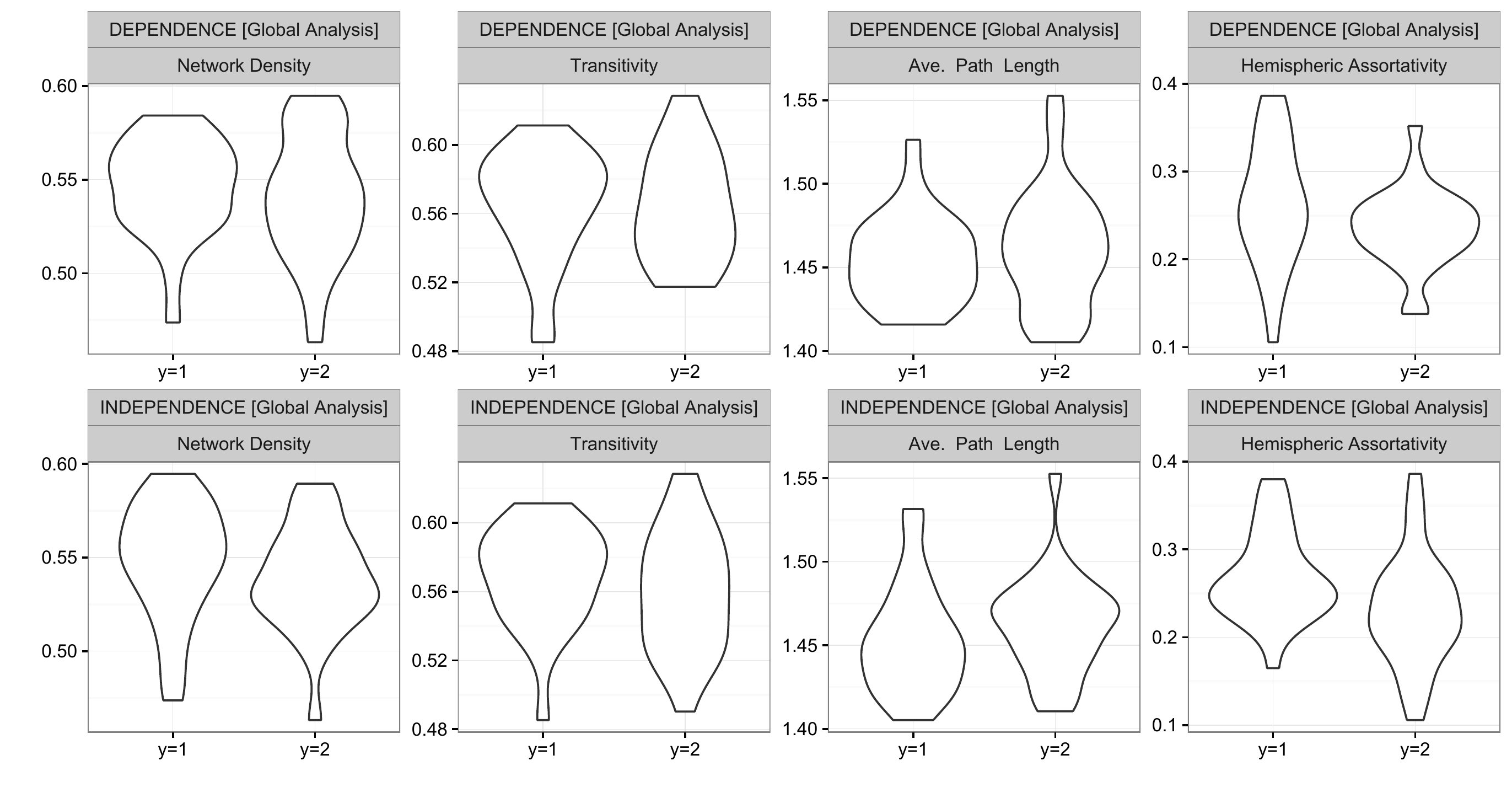}
\caption{{\small{For the two scenarios, observed changes across the two groups for selected network summary statistics. These measures are computed for each simulated network under the two scenarios and summarized via violin plots.}}}
\label{F2}
\end{figure}
\begin{figure}[h!]
\centering
\includegraphics[trim=0.8cm 0.8cm 0.65cm 0.1cm, clip=true,width=11cm]{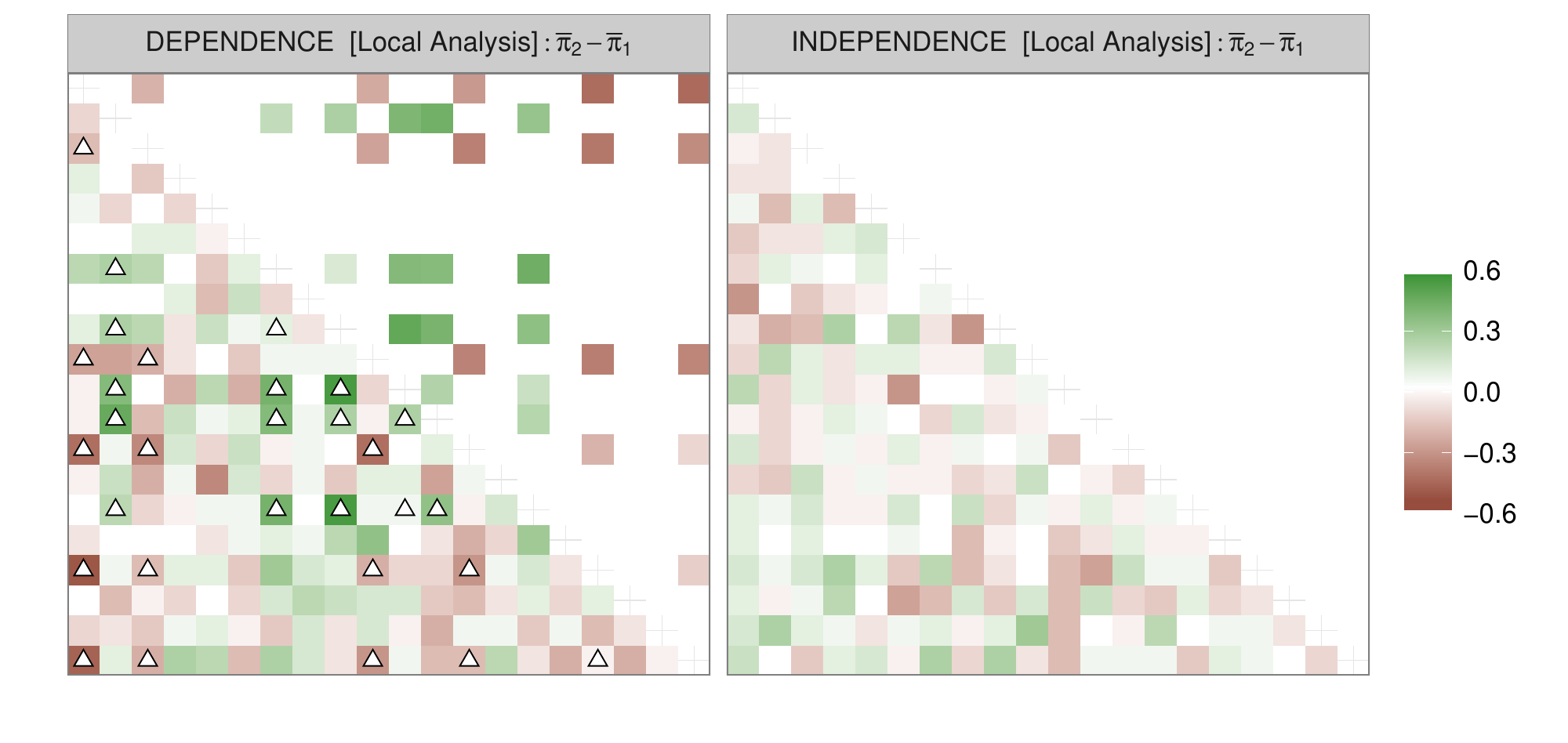}
\caption{{\small{Lower triangular: Group difference between the empirical edge probabilities for each pair of nodes computed from the simulated data. Upper triangular: True group difference in the edge probabilities arising from the generative processes considered in the simulations. These quantities are displayed for the dependence (left) and independence (right) scenarios. Triangles highlight edge probabilities which truly differ across groups in the dependence scenario.}}}
\label{F3}
\end{figure}

\subsection{Simulation settings}
We simulate $n=50$ pairs $(y_i,\boldsymbol{A}_i)$ from our model \eqref{eq1} and \eqref{eq2}--\eqref{eq2_1}, with $y_i$ from a categorical random variable having two equally likely groups ${p}^0_{\mathcal{Y}}(1)=p^0_{\mathcal{Y}}(2)=0.5$ and $\boldsymbol{A}_i$ a $V \times V$ network with $V=20$ nodes.  We consider $H=2$ mixture components, with $\boldsymbol{\pi}^{0(h)}$ defined as in \eqref{eq2_1}. Brain networks are typically characterized by tighter intra-hemispheric than inter-hemispheric connections \citep{ronc_2013}. Hence, we consider two node blocks $V_1=\{1,\ldots,10 \}$ and $V_2=\{11, \ldots, 20 \}$ characterizing left and right hemisphere, respectively, and generate entries in $\boldsymbol{Z}^0$ to favor more likely connections between pairs in the same block than pairs in different blocks. 

To assess performance in local testing, we induce group differences in the connections for a small subset of nodes $V^*\subset \{1, \ldots, V\}$. To include this scenario we consider $R=1$, $\lambda_1^{0(1)}=\lambda_1^{0(2)}=1$ and let $X_{v1}^{0(h)} \neq 0$ only for nodes $v \in V^*$, while fixing the latent coordinates of the remaining nodes to $0$. As a result,  no variations in edge probabilities are displayed when the mixing probabilities remain constant, while only local differences are highlighted when the mixing probabilities shift across groups. Under the dependence scenario, data are simulated with group-specific mixing probabilities $\boldsymbol{\nu}_1^0=(0.8,0.2)$, $\boldsymbol{\nu}_2^0=(0.2,0.8)$. Instead, equal mixing probabilities $\boldsymbol{\nu}_1^0=\boldsymbol{\nu}_2^0=(0.5,0.5)$ are considered under independence. Although we focus on only  $V=20$ nodes to facilitate graphical analyses, the mixture representation in \eqref{eq2} and the low-rank factorization in \eqref{eq2_1} allows scaling to higher $V$ settings. 

As shown in Figures \ref{F2}--\ref{F3}, although our dependence simulation scenario may appear --- at first --- simple, it provides a challenging setting for procedures assessing evidence of global association by testing on variations in the network summary measures. In fact, we choose values $X_{v1}^{0(h)}$ for the nodes $v \in V^*$ such that the resulting summary statistics for the simulated networks do not display changes across groups also in the dependence scenario. Hence, a global test relying on network summary measures is expected to fail in detecting association between $\mathcal{Y}$ and $\mathcal{L}(\boldsymbol{\mathcal{A}})$, as variations in the networks' pmf are only local --- i.e. in a subset of its marginals $\mathcal{L}({\mathcal{A}})_l$. On the other hand, powerful local testing procedures are required to efficiently detect this small set of edge probabilities truly changing across the two groups.

\begin{figure}[t]
\centering
\includegraphics[trim=0.65cm 0.75cm 0.65cm 0cm, clip=true,height=4.5cm, width=13.5cm]{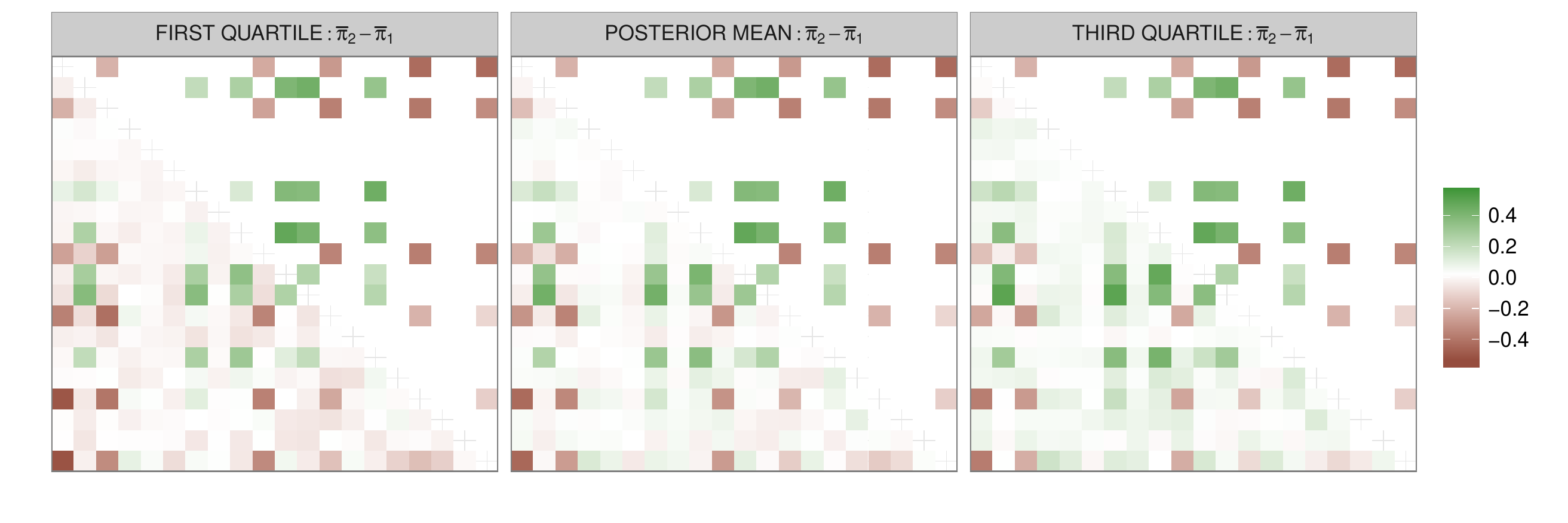}
\caption{{\small{Lower triangular: For the dependence simulation scenario, mean and quartiles of the posterior distribution for the difference between the edge probabilities in the second group ${\bar{\pi}}_{2l}$ and first group ${\bar{\pi}}_{1l}$, $l=1, \ldots, V(V-1)/2$. Upper triangular: For the same scenario, true difference ${\bar{\pi}}^{0}_{2l}-{\bar{\pi}}^{0}_{1l}$, $l=1, \ldots, V(V-1)/2$. In the Figure, the pairs of nodes --- indexed by $l$ --- are re-arranged in matrix form.}}}
\label{F5}
\end{figure}

In both scenarios, inference is accomplished by considering $H=R=10$, $\mbox{pr}(H_1)=\mbox{pr}(H_0)=0.5$ and letting $1-p_{\mathcal{Y}}(2)=p_{\mathcal{Y}}(1) \sim \mbox{Beta}(1/2,1/2)$. For priors $\Pi_Z, \Pi_X$ and $\Pi_{\lambda}$, we choose the same default hyperparameters suggested by \citet{dur_2014}. We collect $5{,}000$ Gibbs iterations, discarding the first $1{,}000$. In both scenarios convergence and mixing are assessed via \citet{gelm_1992} potential scale reduction factors (PSRF) and effective sample sizes, respectively. The PSRFs are obtained by splitting each chain in four consecutive sub-chains of length $1{,}000$ after burn-in, and comparing between and within sub-chains variance. Convergence and mixing assessments focus on parameters of interest for inference, including the Cramer's V coefficients $\rho_l$, $l=1, \ldots, V(V-1)/2$ for local testing and the group-specific edge probability vectors $\boldsymbol{\bar{\pi}}_{y}$, with elements $\bar{\pi}_{yl}=p_{\mathcal{L}({\mathcal{A}})_l \mid y}(1)=\mbox{pr}\{\mathcal{L}(\mathcal{A})_l=1 \mid \mathcal{Y}=y \}$ defined in Proposition \ref{lem2.1}.  This vector coincides with the group-specific mean network structure $\mbox{E}\{\mathcal{L}(\boldsymbol{\mathcal{A}}) \mid \mathcal{Y}=y\}=\sum_{\boldsymbol{a} \in \mathbb{A}_V}\boldsymbol{a} \times p_{\mathcal{L}(\boldsymbol{\mathcal{A}}) \mid y}(\boldsymbol{a})=\sum_{h=1}^{H}\nu_{hy}\boldsymbol{\pi}^{(h)}$ under factorization (\ref{eq2}). In both scenarios, most of the effective samples sizes are around $2{,}000$ out of $4{,}000$ samples, demonstrating excellent mixing performance. Similarly, all the PSRFs are less than $1.1$, providing evidence that convergence has been reached.

\subsection{Global and local testing performance}
Our testing procedure allows accurate inference on the global association between $\mathcal{L}(\boldsymbol{\mathcal{A}})$ and $\mathcal{Y}$. We obtain $\hat{\mbox{pr}}[H_1 \mid \{\boldsymbol{y}, \mathcal{L}(\boldsymbol{A})\}]>0.99$ for the dependence scenario and $\hat{\mbox{pr}}[H_1 \mid \{\boldsymbol{y}, \mathcal{L}(\boldsymbol{A})\}]<0.01$ when $y_i$ and $\boldsymbol{A}_i$, $i=1, \ldots, n$ are generated independently. Instead, the MANOVA testing procedure on the summary statistics vector fails to reject the null hypothesis of no association in both scenarios at a level $\alpha=0.1$ --- as expected. This result further highlights how global network measures may fail in accurately characterizing the whole network architecture.

\begin{figure}[t]
\centering
\includegraphics[trim=0.8cm 0.8cm 0.65cm 0.1cm, clip=true, width=11cm]{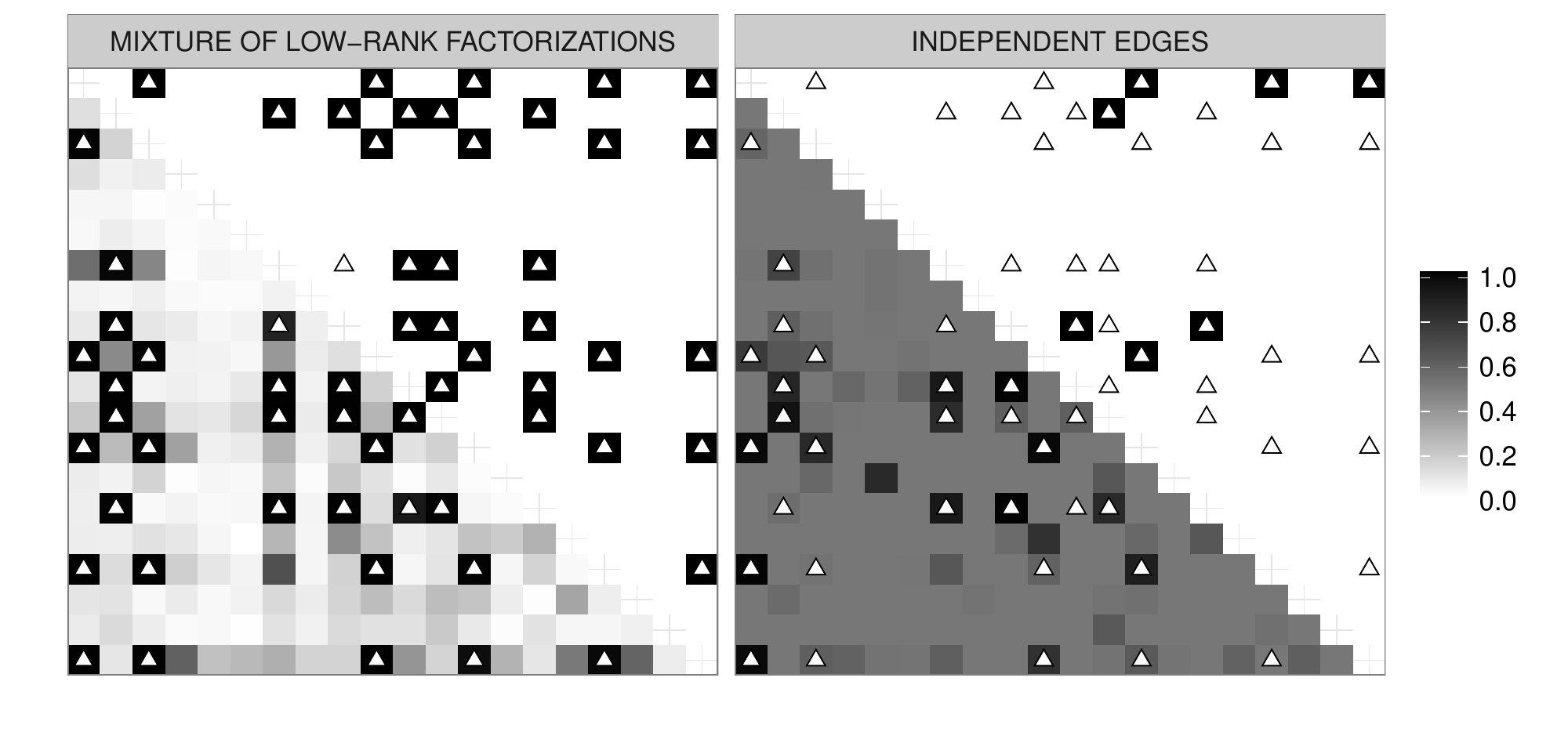}
\vspace{-0.5\baselineskip}
\caption{{\small{Lower triangular: $\hat{\mbox{pr}}[H_{1l} \mid \{\boldsymbol{y}, \mathcal{L}(\boldsymbol{A})\}]=\hat{\mbox{pr}}[\rho_{l}>0.1\mid \{\boldsymbol{y}, \mathcal{L}(\boldsymbol{A})\}]$ (left) and calibrated  Fisher's exact tests $p$-values $1/(1-e p_l \log p_l)$ if $p_l <1/e$, $0.5$ otherwise (right), to allow comparison with $\hat{\mbox{pr}}[H_{1l} \mid \{\boldsymbol{y}, \mathcal{L}(\boldsymbol{A})\}]$, for each $l=1, \ldots, V(V-1)/2$. Upper triangular: Rejected local null hypotheses (black). Triangles highlight edge probabilities which truly differ across groups. In the Figure, the pairs of nodes --- indexed by $l$ --- are re-arranged in matrix form.}}}
\label{F6}
\end{figure}

Focusing on local testing in the dependence scenario, Figure \ref{F5} shows how accounting for sparsity and network information --- via our dependent mixture of low-rank factorizations --- provides accurate inference on local variations in edge probabilities, correctly highlighting pairs of nodes whose connectivity differs across groups and explicitly characterizing uncertainty through the posterior distribution. Conducting inference on each pair of nodes separately provides instead poor estimates --- refer to left plot in Figure \ref{F3} --- with the sub-optimality arising from inefficient borrowing of information across the edges. This lack of efficiency strongly affects also the local testing performance as shown in Figure \ref{F6}, with our procedure having higher power than the one obtained via separate Fisher's exact tests. In Figure  \ref{F6}, each Fisher's exact test $p$-value is calibrated via $1/(1-e p_l \log p_l)$ if $p_l <1/e$ and $0.5$ otherwise, to allow better comparison with $\hat{\mbox{pr}}[H_{1l} \mid \{\boldsymbol{y}, \mathcal{L}(\boldsymbol{A})\}]$  \citep{sel_2001}. Moreover, we adjust for multiplicity in the Fisher's exact tests by rejecting all the local nulls having a $p$-value below $p^*$, with $p^*$ the \citet{ben_1995} threshold to maintain a false discovery rate FDR $\leq 0.1$. Under our local Bayesian testing procedure we reject all $H_{0l}$ such that $\hat{\mbox{pr}}[H_{1l} \mid \{\boldsymbol{y}, \mathcal{L}(\boldsymbol{A})\}]> 0.9$, with $\epsilon=0.1$. We do not explicitly control for FDR in order to assess whether our Bayesian procedures contain the intrinsic adjustment for multiple testing we expect. According to Figure \ref{F6}, thresholding the posterior probability of the local alternatives allows implicit adjustment for multiple testings.  When explicit FDR control is required, one possibility is to define the threshold following the notion of Bayesian false discovery rate in \cite{Newton_2004}.

To assess frequentist operating characteristics, we repeated the above simulation exercise for 100 simulated datasets under both dependence and independence scenarios.  The MANOVA test is performed under a threshold $\alpha=0.1$, while the decision rule in the local Fisher's exact tests is based on the \citet{ben_1995} threshold to maintain a false discovery rate FDR $\leq 0.1$. Under our Bayesian procedure we reject the global null if  $\hat{\mbox{pr}}[H_{1} \mid \{\boldsymbol{y}, \mathcal{L}(\boldsymbol{A})\}]>0.9$. As the prior odds are $\mbox{pr}(H_1)/\mbox{pr}(H_0)=1$, the chosen value $0.9$ implies a threshold on the Bayes factor for significance close to the strong evidence bar suggested by  \citet{kass_1995}. According to sensitivity analyses, moderate changes in the threshold do not affect the final conclusions. Consistently with our initial simulation, we reject local nulls if $\hat{\mbox{pr}}[H_{1l} \mid \{\boldsymbol{y}, \mathcal{L}(\boldsymbol{A})\}]>0.9$.  Also in this case results are not substantially affected by moderate changes in the threshold both in simulation and application; hence, we maintain this choice to preserve coherence in our analyses. 

\begin{table}[t]
		\centering
		\label{tab:1}

				\begin{tabular}{lcccc}

&Type I error & Type II error &FWER &FDR\\
		\hline
&\multicolumn{4}{c}{Global testing procedure}\\
\hline
Mixture of low-rank factorizations&$0.01$&$0.01$& &\\
MANOVA on summary measures&$0.09$&$0.90$& &\\ 
\hline
&\multicolumn{4}{c}{Local testing procedure}\\
\hline
Mixture of low-rank factorizations&$0.0004$&$0.0587$& $0.0600$&$0.0023$\\
Separate Fisher's exact tests&$0.0036$&$0.5983$&$0.4000$&$0.0387$\\ 
			\hline
			\end{tabular}
\caption{{\small{Comparison of error rates for our procedure against MANOVA on summary statistics for global testing and separate Fisher's exact tests for local hypotheses.}}}
\label{tab:1}

	\end{table}
\begin{table}[t]
		\centering
		\label{tab:2}

				\begin{tabular}{lcccc}

&Minimum & Mean &Median &Maximum\\
		\hline
&\multicolumn{4}{c}{Area under the ROC curve (AUC)}\\
\hline
Mixture of low-rank factorizations&$0.969$&$0.999$& $1.000$&$1.000$\\
Separate Fisher's exact tests&$0.810$&$0.921$&$0.923$&$0.989$\\ 
			\hline
			\end{tabular}
\caption{{\small{Summary of the AUCs computed for the $100$ simulated datasets in the dependence scenario, to assess performance of local testing at varying thresholds. The ROC curves are constructed using the true hypotheses indicators --- $\delta_l=0$ if $H_{0l}$ is true, $\delta_l=1$ if $H_{1l}$ is true, $l=1, \ldots, V(V-1)/2$ --- and the acceptance or rejection decisions based on our procedure and Fisher's exact tests at varying the thresholds on posterior probabilities or FDR, respectively.}}}
\label{tab:2}

	\end{table}

\begin{figure}[t]
\centering
\includegraphics[trim=0.8cm 0.1cm 0.65cm 0.1cm, clip=true,height=4.8cm, width=13.5cm]{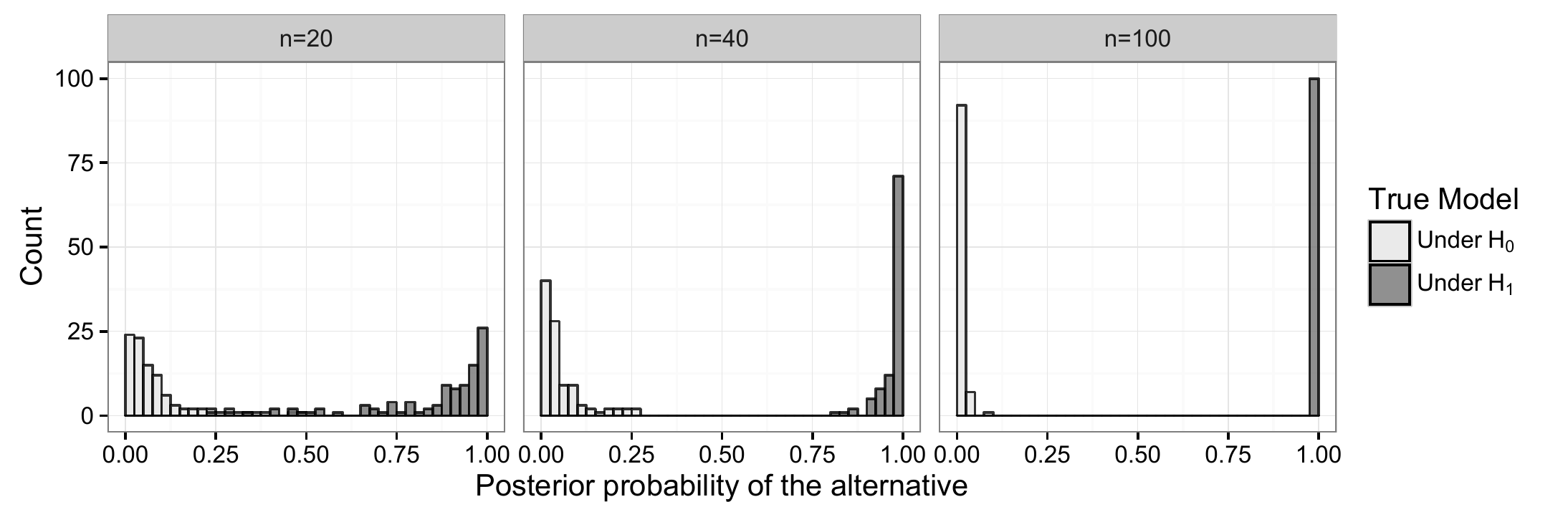}
\caption{{\small{For increasing sample sizes $n$, histograms of the estimated posterior probabilities of the global alternative $H_1$ in each of the 100 simulations under dependence and independence.}}}
\label{F7}
\end{figure}

Table \ref{tab:1} confirms the superior performance of our approach in maintaining all error rates close to zero, in both global and local testing, while intrinsically adjusting for multiplicity. 
The information reduction via summary measures for the global test and the lack of a network structure in the local Fisher's exact tests lead to procedures with substantially less power. Although Table \ref{tab:1} has been constructed using an FDR control of $0.1$ in the Fisher's exact tests and a threshold of $0.9$ under our local testing procedure, we maintain superior performance allowing the thresholds to vary, as shown in Table  \ref{tab:2}.

\begin{figure}[t]
\centering
\includegraphics[trim=0.2cm 0.1cm 0.1cm 0cm, clip=true,height=8cm, width=13.5cm]{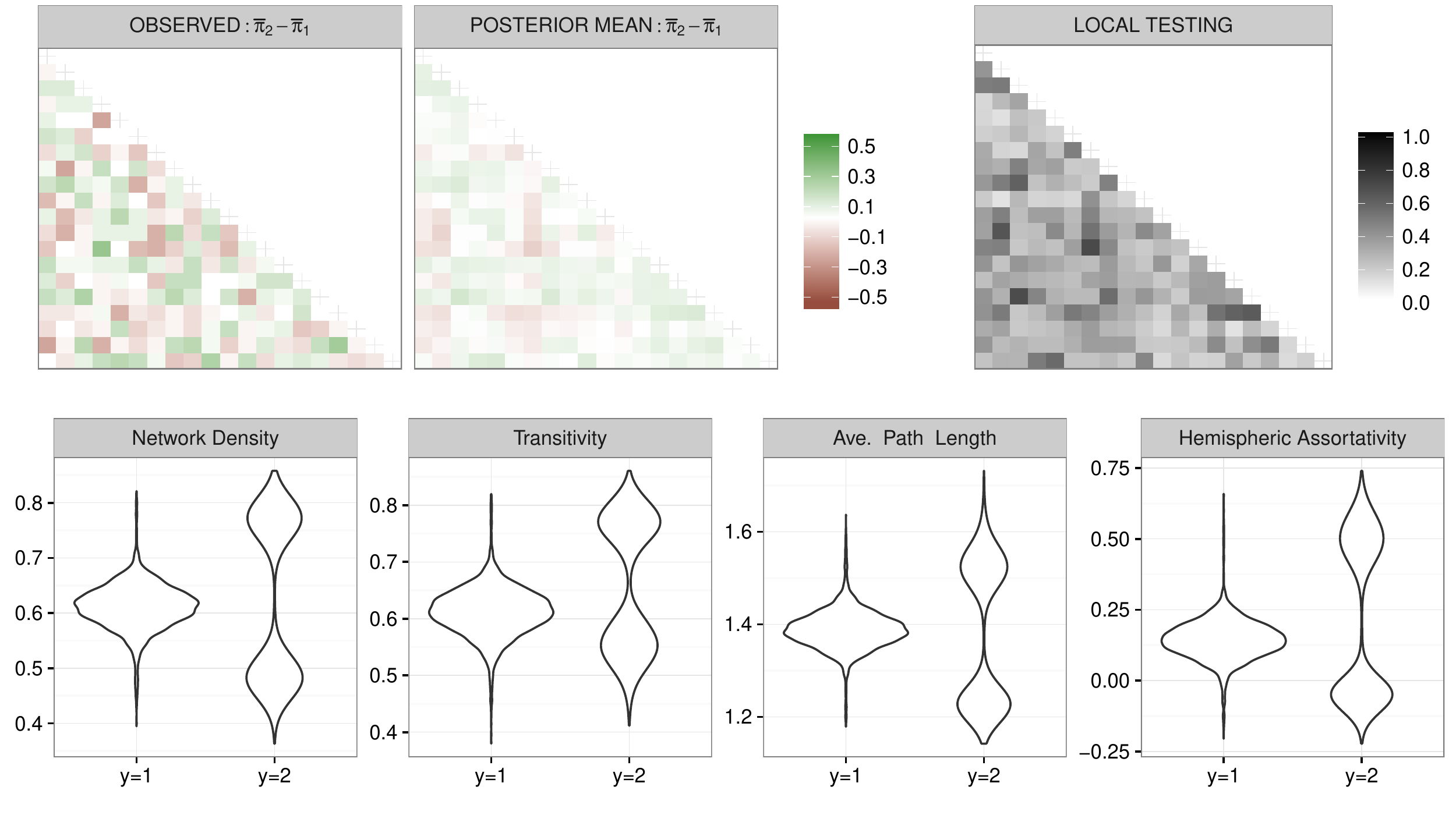}
\caption{{\small{Model performance in the final simulation scenario. Upper-left matrix: Group difference between the empirical edge probabilities for each pair of nodes computed from the simulated data (lower triangular) versus the true group difference in the edge probabilities  (upper triangular). Upper-middle matrix: Posterior mean of the difference between the edge probabilities in the two groups (lower triangular) versus true group difference in edge probabilities  (upper triangular). Upper-right matrix: $\hat{\mbox{pr}}[H_{1l} \mid \{\boldsymbol{y}, \mathcal{L}(\boldsymbol{A})\}]=\hat{\mbox{pr}}[\rho_{l}>0.1\mid \{\boldsymbol{y}, \mathcal{L}(\boldsymbol{A})\}]$ (lower triangular) and  rejected (black) local null hypotheses (upper triangular), for $l=1, \ldots, V(V-1)/2$ --- re-arranged in matrix form. Lower panels: Violin plots representing the posterior predictive distribution of selected network summary statistics in the two groups, arising from our model.}}}
\label{F8}
\end{figure}

In considering sample size versus type I and type II error rates, it is interesting to assess the rate at which the posterior probability of the global alternative
${\mbox{pr}}[H_{1} \mid \{\boldsymbol{y}, \mathcal{L}(\boldsymbol{A})\}]$ converges to 0 and 1 under $H_0$ and $H_1$, respectively, as $n$ increases. We evaluate this behavior by simulating $100$ datasets as in the previous simulation for increasing sample sizes $n=20$, $n=40$ and $n=100$ and for each scenario. Figure \ref{F7} provides histograms showing the estimated posterior probabilities of $H_1$ for the 100 simulated datasets under the two scenarios and for increasing sample sizes. The separation between  scenarios is evident for all sample sizes, with $\hat{\mbox{pr}}[H_{1} \mid \{\boldsymbol{y}, \mathcal{L}(\boldsymbol{A})\}]$ consistently concentrating close to $0$ and $1$ under the independence and dependence scenario, respectively, as $n$ increases. When $n=20$ the test has lower power, with $32/100$ samples having  $\hat{\mbox{pr}}[H_{1} \mid \{\boldsymbol{y}, \mathcal{L}(\boldsymbol{A})\}]<0.9$ when $H_1$ is true.  However, type I errors were rare, with $1/100$ samples having $\hat{\mbox{pr}}[H_{1} \mid \{\boldsymbol{y}, \mathcal{L}(\boldsymbol{A})\}]>0.9$ when data are generated under $H_0$. These values are  very close to 0 when the sample size is increased to $n=40$ and $n=100$, with the latter showing strongly concentrated estimates close to $0$ and $1$, when $H_0$ is true and $H_1$ is true, respectively.

\subsection{Identifying group differences in more complex functionals}
We conclude our simulation studies by considering a scenario in which there is a strong dependence between $\mathcal{L}(\boldsymbol{\mathcal{A}})$ and $\mathcal{Y}$, but this dependence arises from changes in more complex structures, instead of just the edge probabilities. Specifically, we simulate $n=50$ pairs $(y_i,\boldsymbol{A}_i)$ from our model \eqref{eq1} and \eqref{eq2}, with ${p}^0_{\mathcal{Y}}(1)=p^0_{\mathcal{Y}}(2)=0.5$ and $\boldsymbol{A}_i$ a $V \times V$ network with $V=20$ nodes.  In defining \eqref{eq2} we consider $H=3$ components and again split the nodes in two blocks $V_1=\{1, \ldots, 10 \}$ and $V_2=\{11, \ldots, 20 \}$, characterizing --- for example --- the two different hemispheres. When $h=1$, the vector $\boldsymbol{\pi}^{0(1)}$ characterizes this block structure, with the probability of an edge between pairs of nodes in the same block set at $0.75$, while nodes in different blocks have $0.5$ probability to be connected. Vectors $\boldsymbol{\pi}^{0(2)}$ and $\boldsymbol{\pi}^{0(3)}$ maintain the same within block probability of $0.75$ as in $\boldsymbol{\pi}^{0(1)}$, but have different across block probability. In component $h=2$ the latter increases by $0.3$ --- from $0.5$ to $0.8$ --- while in component $h=3$ this quantity decreases by the same value --- from $0.5$ to $0.2$. As a result, when letting $\boldsymbol{\nu}_1^0=(1,0,0)$ and $\boldsymbol{\nu}_2^0=(0,0.5,0.5)$ it is easy to show that the group-specific edge probabilities --- characterizing the distribution of each edge in the two groups --- remain equal  $\boldsymbol{\bar{\pi}}^0_{1}=\boldsymbol{\bar{\pi}}^0_{2}$, even if the probability mass function jointly assigned to these edges changes across groups ${p}^0_{\mathcal{L}(\boldsymbol{\mathcal{A}}) \mid 1}\neq {p}^0_{\mathcal{L}(\boldsymbol{\mathcal{A}}) \mid 2}$. 

This provide a subtle scenario for the several procedures assessing evidence of changes in the brain network across groups, by focusing solely on marginal or expected quantities. These strategies should --- correctly --- find no difference in edge probabilities and hence may be --- wrongly --- prone to conclude that the brain network does not change across groups. Underestimating associations may be a dangerous fallacy in understating --- for example --- the effect of a neurological disorder that  induces changes in more complex functionals of the brain network.

We apply our procedures to these simulated data under the same settings of our initial simulations, obtaining very similar effective sample sizes and PSRFs. As shown in the upper panels of Figure \ref{F8} the posterior probabilities for all the local alternatives are lower than $0.9$ and hence our multiple testing procedure does not reject $H_{0l}$ for every $l=1, \ldots, V(V-1)/2$. Beside correctly assessing the evidence of no changes in edge probabilities across the two groups, our global test is able to detect variations in more complex functionals of the brain network. In fact we obtain  $\hat{\mbox{pr}}[H_{1} \mid \{\boldsymbol{y}, \mathcal{L}(\boldsymbol{A})\}]>0.99$, meaning that although  there is no evidence of changes in edge probabilities across the two groups, the model finds a strong association between $\mathcal{L}(\boldsymbol{\mathcal{A}})$ and $\mathcal{Y}$. 

The type of variations in more complex structures can be observed in the lower panels of Figure \ref{F8} showing the posterior predictive distribution  of the selected network summary statistics obtained under our statistical model. Although the latter is not analytically available, it is straightforward to simulate from the posterior predictive distribution exploiting our constructive representation in Figure \ref{F_dep} and posterior samples for the quantities in \eqref{eq1} and \eqref{eq2}--\eqref{eq2_1}. Specifically, for each MCMC sample  of the parameters in \eqref{eq1} and \eqref{eq2}--\eqref{eq2_1} --- after convergence ---  we generate a network from our model exploiting the mechanism in Figure \ref{F_dep}, to obtain the desired samples from  the posterior predictive distribution. According to the lower panels of Figure \ref{F8} there are substantial changes in the pmf of the network data across groups. In group one our model infers network summary measures having unimodal distributions, while in the second group we learn  substantially different bimodal distributions. This behavior was expected based on our simulation, and hence these results further confirm the accuracy of our global test along with  the good performance  of our model in flexibly characterizing the distribution of a network-valued random variable and its variations across groups. 

\section{Application to human brain networks and creativity}
\label{sec:app}
We apply our method to the dataset described in the introduction using the same settings as in the simulation examples, but with upper bound $H$ increased to $H=15$. This choice proves to be sufficient with components $h=12, \ldots, 15$ having no observations and redundant dimensions of the latent spaces efficiently removed. The efficiency of the Gibbs sampler was very good, with effective sample sizes  around $1{,}500$ out of $4{,}000$. Similarly the PSRFs provide evidence that convergence has been reached, as the highest of these quantities is $1.15$. These checks on mixing and convergence are performed for the chains associated with quantities of interest for inference and testing. These include the Cramer's V coefficients $\rho_l$, $l=1, \ldots, V(V-1)/2$, the group-specific edge probability vectors $\boldsymbol{\bar{\pi}}_{1}$, $\boldsymbol{\bar{\pi}}_{2}$ and the expectation of selected network summary statistics. 

Our results provide interesting insights into the global relation between the brain network and creativity, with $\hat{\mbox{pr}}[H_{1} \mid \{\boldsymbol{y}, \mathcal{L}(\boldsymbol{A})\}]=0.995$ strongly favoring the alternative hypothesis of association between the brain connectivity architecture and the level of creative reasoning.  To assess the robustness of our global test, we also performed posterior computation based on datasets that randomly matched the observed group membership variables with a corresponding brain network, effectively removing the possibility of an association. In $10$ of these trials we always obtained --- as expected ---  low $\hat{\mbox{pr}}[H_{1} \mid \{\boldsymbol{y}, \mathcal{L}(\boldsymbol{A})\}] \leq 0.2$. 

We also attempted to apply the MANOVA test as implemented in the simulation experiments, with the same network statistics --- i.e. network density, transitivity, average path length and assortativity by hemisphere. These are popular measures in neuroscience in informing on fundamental properties in brain network organization, such as small-world, homophily patterns and scale-free behaviors \citep{Bull_2009,Rub_2010,bull_2012}. In our dataset, the average path length was undefined for three subjects, as there were no paths between several pairs of their brain regions.  Replacing these undefined shortest path lengths with the maximum path length, we observe no significant changes across creativity groups with a $p$-value of $0.111$.  When excluding this topological measure, we obtain a borderline  $p$-value of $0.054$.  This sensitivity to the choice of summary statistics further motivates tests that avoid choosing topological measures, which is a somewhat arbitrary exercise.

\begin{figure}[t]
\centering
\includegraphics[trim=0.65cm 0.75cm 0.65cm 0cm, clip=true, width=13cm]{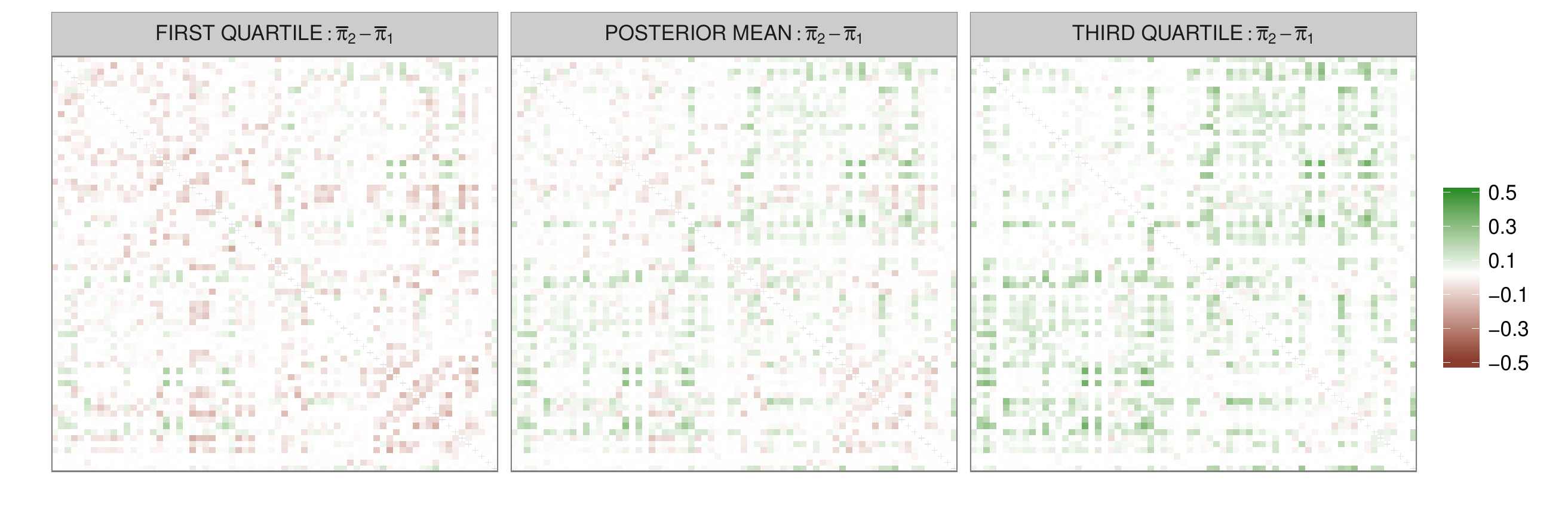}
\caption{{\small{Mean and quartiles of the posterior distribution for the difference ${\bar{\pi}}_{2l}-{\bar{\pi}}_{1l}$ between the edge probabilities in high  and low creativity subjects, for each pair $l=1, \ldots, V(V-1)/2$.  In the Figure, the pairs of brain regions --- indexed by $l$ --- are re-arranged in matrix form.}}}
\label{F9}
\end{figure}

\begin{figure}[t]
\centering
\includegraphics[trim=0.65cm 0.2cm 0cm 0cm, clip=true,height=11cm, width=14cm]{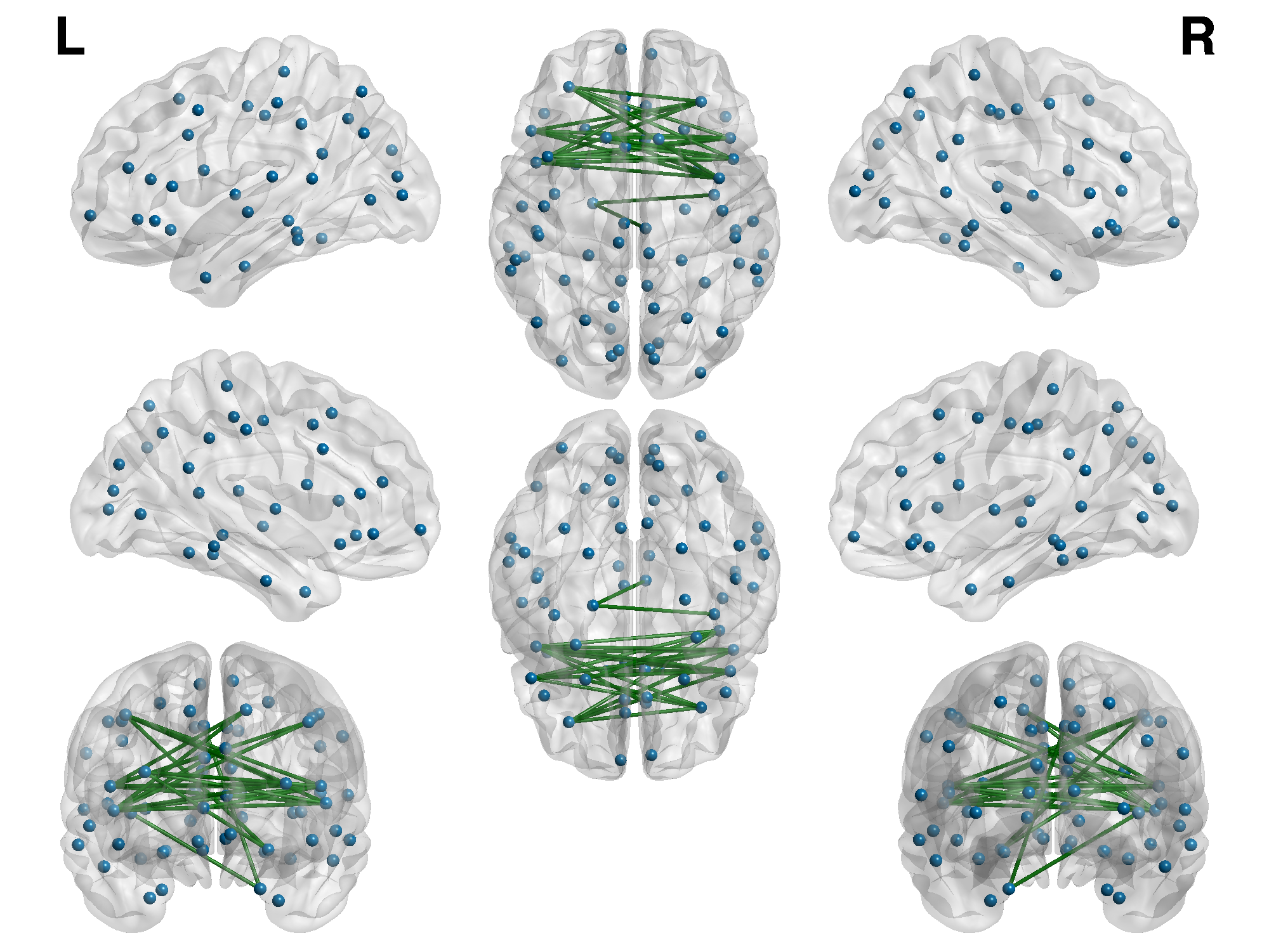}
\caption[Results from local testing in the creativity application]{\footnotesize{Brain network visualization exploiting results from our local testing procedure. We only display those connections which provide evidence of changes across high and low creativity subjects based on our procedure -- i.e. $\hat{\mbox{pr}}[H_{1l} \mid \{\boldsymbol{y}, \mathcal{L}(\boldsymbol{A})\}]>0.9$. Edge color is green -- or red -- if its estimated probability in high creativity subjects is greater -- or less --  than low creativity ones.  Regions' positions are given by their spatial coordinates in the brain, with the same brain displayed from different views.}}
\label{F10}
\end{figure}

\begin{figure}[t]
\centering
\includegraphics[trim=0.75cm 0cm 0.22cm 0cm, clip=true,width=11cm]{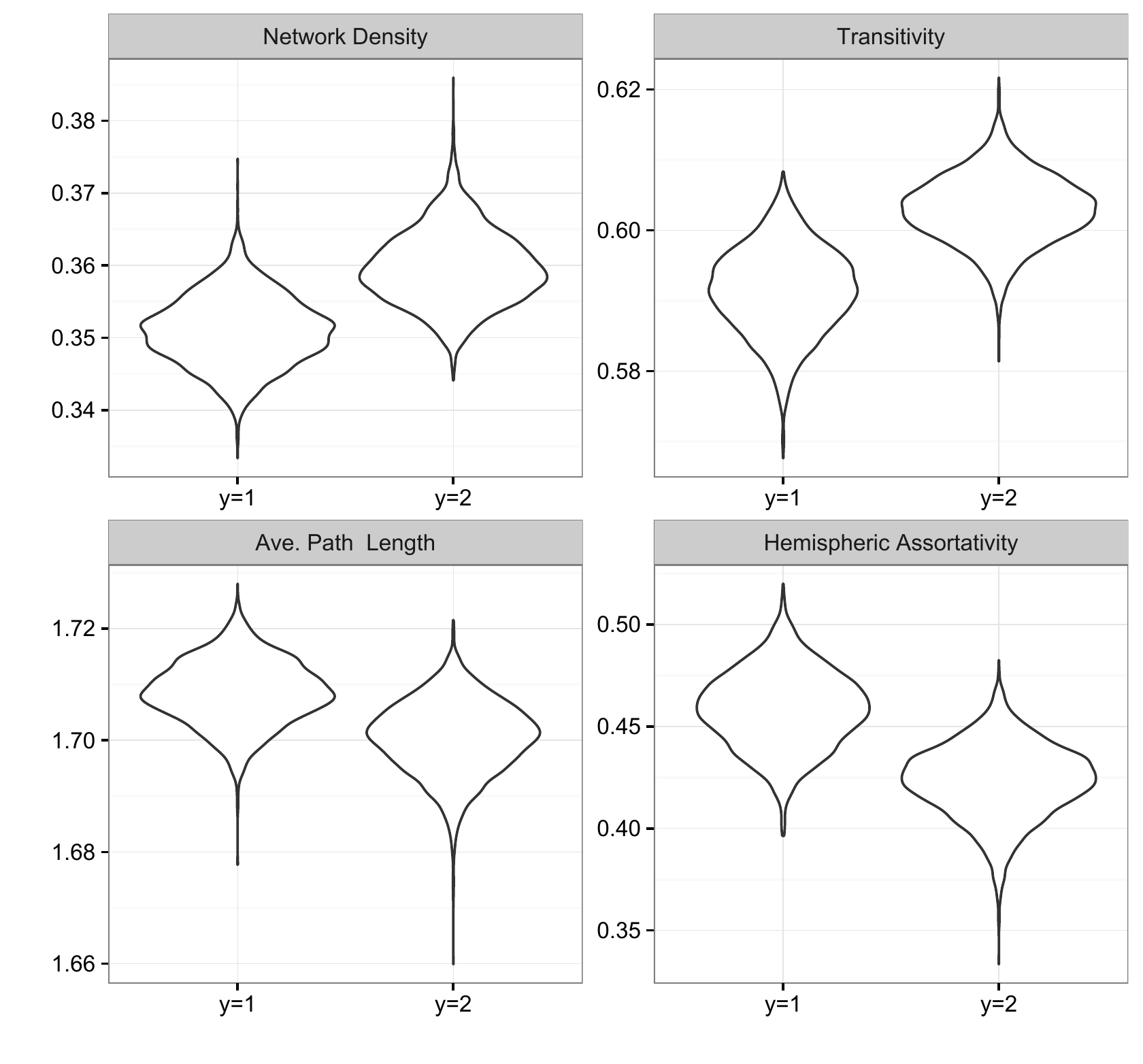}
\caption{{\small{Violin plots representing the posterior distribution for the expectation of selected network summary statistics in the two creativity groups.}}}
\label{F12}
\end{figure}


As a secondary focus, we also examined predictive performance of our model.  In particular, we considered in-sample edge prediction based on the posterior mean of the edge probabilities in the two groups.  This produced excellent results, with an  area under the ROC curve (AUC) equal to $0.97$. The ROC curve is constructed using the observed edges $\mathcal{L}({A}_{i})_l$, $i=1, \ldots, n$, $l=1, \ldots, V(V-1)/2$ and those predicted with the posterior mean of the group-specific edge probabilities at varying thresholds --- using ${\hat{\bar{\pi}}}_{1l}$ for subjects with $y_i=1$ and ${\hat{\bar{\pi}}}_{2l}$ for subjects with $y_i=2$. 

Beside providing a flexible approach for joint modeling of networks and categorical traits, our model also represents a powerful tool to predict $y_i$ given the subject's full brain network structure. In fact, under our formulation, the probability that a subject $i$ has high creativity, conditionally on his brain structural connectivity network $\boldsymbol{A}_i$, is 
\begin{eqnarray}
\mbox{pr}\{\mathcal{Y}_i = 2 \mid \mathcal{L}(\boldsymbol{A}_{i})\} = 1-\mbox{pr}\{\mathcal{Y}_i =1 \mid \mathcal{L}(\boldsymbol{A}_{i})\} =\frac{p_{\mathcal{Y}}(2)p_{\mathcal{L}(\boldsymbol{\mathcal{A}}) \mid 2}(\boldsymbol{a}_{i})}{p_{\mathcal{Y}}(2)p_{\mathcal{L}(\boldsymbol{\mathcal{A}}) \mid 2}(\boldsymbol{a}_{i})+p_{\mathcal{Y}}(1)p_{\mathcal{L}(\boldsymbol{\mathcal{A}}) \mid 1}(\boldsymbol{a}_{i})}, \nonumber
\end{eqnarray}
where $\boldsymbol{a}_{i}= \mathcal{L}(\boldsymbol{A}_{i})$ is the network configuration of the $i$th subject and $p_{\mathcal{L}(\boldsymbol{\mathcal{A}}) \mid y}(\boldsymbol{a}_{i})$, $y \in\{1,2\}$ can be easily computed from \eqref{eq2}. We obtain an in-sample $\mbox{AUC}=0.87$ in predicting the creativity group $y_i$ using the posterior mean of $\mbox{pr}\{\mathcal{Y}_i = 2 \mid \mathcal{L}(\boldsymbol{A}_{i})\}=1-\mbox{pr}\{\mathcal{Y}_i =1 \mid \mathcal{L}(\boldsymbol{A}_{i})\}$ for each $i=1, \ldots, n$. Hence,  allowing the conditional pmf of the network-valued random variable to shift across groups via group-specific mixing probabilities provides a good characterization of the relation between brains and creativity, leading to accurate prediction of the creativity group.  Although these results are in-sample, they provide reassurance that the substantial dimensionality reduction underlying our representation does not lead to inadequate fit.

Figure \ref{F9} provides summaries of the posterior distribution for the quantities in $\boldsymbol{\bar{\pi}}_{2}-\boldsymbol{\bar{\pi}}_{1}$, with  $\boldsymbol{\bar{\pi}}_{2}=\sum_{h=1}^{H}\nu_{h2}\boldsymbol{\pi}^{(h)}$ and $\boldsymbol{\bar{\pi}}_{1}=\sum_{h=1}^{H}\nu_{h1}\boldsymbol{\pi}^{(h)}$ encoding the edge probabilities in high and low creativity groups, respectively. Most of these connections have a similar probability in the two groups, with more evident local differences for connections among brain regions in different hemispheres. Highly creative individuals display a higher propensity to form inter-hemispheric connections. Differences in intra-hemispheric circuits are less evident. These findings are confirmed by Figure \ref{F10} including also results from our local testing procedure. As in the simulation, we set $\epsilon=0.1$ and the decision rule rejects the local null $H_{0l}$ when $\hat{\mbox{pr}}[H_{1l} \mid \{\boldsymbol{y}, \mathcal{L}(\boldsymbol{A})\}]>0.9$. These choices provide reasonable settings based on simulations, and results are robust to moderate changes in the thresholds.

Previous studies show that intra-hemispheric connections are more likely than inter-hemispheric connections for healthy individuals  \citep{ronc_2013}. This is also evident in our dataset, with subjects having a proportion of intra-hemispheric edges of 0.55 over the total number of possible intra-hemispheric connections, against a proportion of about 0.21 for the inter-hemispheric ones. 
Our estimates in Figure \ref{F9} and local tests in Figure \ref{F10} highlight differences only in terms of inter-hemispheric connectivity, with highly creative subjects having a stronger propensity to connect regions in different hemispheres. This is consistent with the idea that creative innovations arise from communication of brain regions that ordinarily are not  connected \citep{hei_2003}.

These findings contribute to the ongoing debate on the sources of creativity in the human brain, with original theories considering the right-hemisphere as the seat of creative thinking, and more recent empirical analyses highlighting the importance of the level of communication between the two hemispheres of the brain; see \citet{saw_2012}, \citet{sho_2009} and the references cited therein. Beside the differences in techniques to monitor brain networks and measure creativity, as stated in \citet{ard_2010}, previous lack of agreement is likely due to the absence of a unifying approach to statistical inference in this field. Our method addresses this issue, while essentially supporting modern theories considering creativity as a result of cooperating hemispheres. 

According to Figure  \ref{F10} the differences in terms of inter-hemispheric connectivity are found mainly in the frontal lobe, where the  co-activation circuits in the high creativity group are denser. This is in line with recent findings highlighting the major role of the frontal lobe in creative cognition \citep{carl_2000,jung_2010,take_2010}. Previous analyses focus on variations in the activity of each region in isolation, with  \citet{carl_2000} and  \citet{take_2010} noticing an increase in cerebral blood flow and fractional anisotropy, respectively, for highly creative subjects, and  \citet{jung_2010} showing a negative association between creativity and cortical thickness in frontal regions. We instead provide inference on the interconnections among these regions, with increased bilateral frontal connectivity for highly creative subjects, consistent with both the attempt to enhance frontal activity as suggested by \citet{carl_2000} and  \citet{take_2010} or reduce it according to \citet{jung_2010}.

Figure  \ref{F12} shows the effect of the increased inter-hemispheric frontal connectivity --- in high creativity subjects --- on the posterior distribution of the key expected network summary statistics in the two groups. Although the expectation for most of these quantities cannot be analytically derived as a function of the parameters in \eqref{eq2}--\eqref{eq2_1}, it is straightforward to obtain posterior samples for the previous measures via Monte Carlo methods exploiting the constructive representation in Figure \ref{F_dep}. According to Figure  \ref{F12} the brains in high creativity subjects are characterized by an improved architecture --- compared to low creativity subjects --- with increased connections, higher transitivity and shortest paths connecting pairs of nodes. As expected also hemispheric assortativity decreases. This is consistent with our local testing procedure providing evidence of increased  inter-hemispheric activity and unchanged intra-hemispheric connectivity structures across the two groups. Previous results are also indicative of small-world structures in highlighting high transitivity and low average path length, with brains for high creativity subjects having a stronger small-world topology than subjects with low creativity. This is a key property in the organization of brain networks  \citep{Bull_2009}.

\section{Discussion}
\label{sec:disc}
This article proposes the first general approach in the literature --- to our knowledge --- for inference and testing of group differences in network-valued data without focusing on pre-specified functionals or reducing the network data to summary statistics prior to inference. The creativity application illustrates substantial benefits of our approach in providing a unifying and powerful methodology to perform inferences on group differences in brain networks, in contrast to current practice, which applies simple statistical tests based on network summary measures or selected functionals.  These tests tend to lack power and be sensitive to the summary statistics and functionals chosen, contributing to the inconsistent results observed in the recent literature.  Although we specifically focus on creativity, our method can be applied in many other settings.  For example, for inferring differences in brain networks with neuropsychiatric diseases. In addition, our approach is applicable to other fields involving network-valued data.  

It is interesting to generalize our procedure to the multiple group case with $y_i \in \{1, \ldots, K\}$. This can be accomplished with minor modifications to the two groups case.  
Specifically, it is sufficient to consider as many mixing probability vectors $\boldsymbol{\nu}_y$ as the total number of groups $K$, replace the beta prior for ${p}_{\mathcal{Y}}$ with a Dirichlet and appropriately modify the Gibbs sampler. Theoretical properties and testing procedures are trivial to extend.  Although generalization to the multiple groups case is straightforward, there may be subtleties in capturing ordering in the changes across many groups.  

There are other interesting ongoing directions. For example, it is important to allow nonparametric shifts in the pmf associated with the network-valued random variable across non-categorical predictor variables, while developing procedures scaling to a number of nodes much larger than $V=68$. Focusing on neuroscience applications,  another important goal is to develop statistical methods that explicitly take into account errors in constructing the brain connection network, including in alignment and in recovering fiber tracts, taking as input the raw imaging data. Our model partially accounts for these errors via the pmfs for the network-valued random variables and the prior distributions for its quantities. However procedures that explicitly account for this noise, may yield improvements in performance, including better uncertainty quantification. 

Finally, it is important to consider generalizations accommodating fiber counts instead of just binary indicators. Incorporating information on weighted edges, data take the form of multivariate counts, again with network-structured dependence. There are subtleties involved in modeling of multivariate counts. It is common to incorporate latent variables in Poisson factor models \citep[e.g][]{duns_2005}. Including this generalization requires minor modifications of our current procedures, however, as noted in \citet{can_2011}, there is a pitfall in such models due to the dual role of the latent variable component in controlling the degree of dependence and the magnitude of over-dispersion in the marginal distributions. \citet{can_2011} address these issues via a rounded kernel method which improves flexibility in modeling count variables. Our current efforts are aimed at adapting these procedures to develop nonparametric approaches for  inference  on the  distribution of weighted networks. 

\section*{Supplementary materials: Proofs of propositions}
The supplementary materials contain proofs of Propositions 2.1, 2.2 and 3.1 providing theoretical support for the methodology developed in the article ``Bayesian Inference and Testing of  Group Differences in Brain Networks''.
\label{sec:intro}
\begin{proof}{\bf{Proposition 2.1 }}
Recalling Lemma 2.1 in \citet{dur_2014} we can always represent the conditional probability $p_{\mathcal{L}(\boldsymbol{\mathcal{A}}) \mid y}(\boldsymbol{a})$ separately for each group $y \in \{1,2\}$ as $$p_{\mathcal{L}(\boldsymbol{\mathcal{A}}) \mid y}(\boldsymbol{a})=\sum_{h=1}^{H_y} \nu^*_{hy} \prod_{l=1}^{V(V-1)/2} (\pi_{l}^{(hy)})^{a_l} (1-  \pi^{(hy)}_{l})^{1-a_l}, \quad \boldsymbol{a} \in \mathbb{A}_V,$$ with each $\pi^{(hy)}_l$ factorized as $\mbox{logit}(\pi^{(hy)}_l)=Z^{(y)}_l+\sum_{r=1}^{R_y}\lambda^{(hy)}_r X^{(hy)}_{vr} X^{(hy)}_{ur}$, $l=1, \ldots, V(V-1)/2$ and $h=1, \ldots, H_y$. Hence Proposition 2.1 follows after choosing $\boldsymbol{\pi}^{(h)}, h=1, \ldots, H$ as the sequence of unique component-specific edge probability vectors $\boldsymbol{\pi}^{(hy)}$ appearing in the above separate factorizations for at least one group $y$, and letting the group-specific mixing probabilities in (2.8) be $\nu_{hy}= \nu^*_{hy}$ if $\boldsymbol{\pi}^{(h)}=\boldsymbol{\pi}^{(hy)}$ and $\nu_{hy}=0$ otherwise.
\end{proof}

\begin{proof}{\bf{Proposition 2.2}}
Recalling factorization (2.8) and letting $\mathbb{A}^{-l}_{V}$ denote the set containing all the possible network configurations for the node pairs except the $l$th one, we have that $p_{\mathcal{L}(\mathcal{A})_l \mid y}(1)$ is equal to
\begin{eqnarray*}
\sum_{\mathbb{A}^{-l}_{V}}\sum_{h=1}^{H} \nu_{hy} \pi_{l}^{(h)}\prod_{l^*\neq l} (\pi_{l^*}^{(h)})^{a_{l^*}} (1-  \pi^{(h)}_{l^*})^{1-a_{l^*}} =\sum_{h=1}^{H} \nu_{hy} \pi_{l}^{(h)}\sum_{\mathbb{A}^{-l}_{V}}\prod_{l^*\neq l} (\pi_{l^*}^{(h)})^{a_{l^*}} (1-  \pi^{(h)}_{l^*})^{1-a_{l^*}} 
\end{eqnarray*}
Then Proposition 2.2 follows after noticing that $\prod_{l^*\neq l} (\pi_{l^*}^{(h)})^{a_{l^*}} (1-  \pi^{(h)}_{l^*})^{1-a_{l^*}}$ is the joint pmf of independent Bernoulli random variables and hence the summation over the joint sample space $\mathbb{A}^{-l}_{V}=\{0,1 \}^{V(V-1)/2-1}$, provides $\sum_{\mathbb{A}^{-l}_{V}}\prod_{l^*\neq l} (\pi_{l^*}^{(h)})^{a_{l^*}} (1-  \pi^{(h)}_{l^*})^{1-a_{l^*}}=1$. 

The proof of $p_{\mathcal{L}(\mathcal{A})_l}(1)=\sum_{y=1}^{2}p_{\mathcal{Y}}(y)\sum_{h=1}^{H}\nu_{hy}\pi^{(h)}_l$ follows directly from the above results after noticing that $p_{\mathcal{L}(\mathcal{A})_l}(1)=\sum_{y=1}^{2}p_{\mathcal{Y},\mathcal{L}(\mathcal{A})_l}(y,1)=\sum_{y=1}^{2}p_{\mathcal{Y}}(y)p_{\mathcal{L}(\mathcal{A})_l \mid y}(1)$.
\end{proof}

\begin{proof}{\bf{Proposition  3.1}}
Recalling the proof of Proposition 2.1 and factorization (2.5) we can always represent $\sum_{y=1}^{2}  \sum_{\boldsymbol{a} \in \mathbb{A}_V} |p_{\mathcal{Y}, \mathcal{L}(\boldsymbol{\mathcal{A}})}(y,\boldsymbol{a})- p^0_{\mathcal{Y}, \mathcal{L}(\boldsymbol{\mathcal{A}})}(y,\boldsymbol{a})|$  as 
{\small{
\begin{eqnarray*}
\sum_{y=1}^2 \sum_{\boldsymbol{a} \in \mathbb{A}_V} |p_{\mathcal{Y}}(y) \sum_{h=1}^{H} \nu_{hy} \prod_{l=1}^{V(V-1)/2} (\pi_{l}^{(h)})^{a_l} (1-  \pi^{(h)}_{l})^{1-a_l} \quad \quad  \\ 
\ \ \ \ \ -p^0_{\mathcal{Y}}(y) \sum_{h=1}^{H} \nu^0_{hy} \prod_{l=1}^{V(V-1)/2} (\pi_{l}^{0(h)})^{a_l} (1-  \pi^{0(h)}_{l})^{1-a_l}|,
\end{eqnarray*}}}with $\nu^0_{hy}=\nu^{*0}_{hy}$ if $\boldsymbol{\pi}^{0(h)}=\boldsymbol{\pi}^{0(hy)}$ and $\nu^0_{hy}=0$ otherwise. Hence $\Pi\{\mathbb{B}_{\epsilon}({p}^0_{\mathcal{Y}, \mathcal{L}(\boldsymbol{\mathcal{A}})})\}$ is 
{\small{
\begin{eqnarray*}
\int 1(\sum_{y=1}^{2}  \sum_{\boldsymbol{a} \in \mathbb{A}_V} |p_{\mathcal{Y}, \mathcal{L}(\boldsymbol{\mathcal{A}})}(y,\boldsymbol{a})- p^0_{\mathcal{Y}, \mathcal{L}(\boldsymbol{\mathcal{A}})}(y,\boldsymbol{a})|<\epsilon) d\Pi_y({p}_{\mathcal{Y}}) d \Pi_{\nu}(\boldsymbol{\nu}_1, \boldsymbol{\nu}_2) d \Pi_{\pi}(\boldsymbol{\pi}^{(1)}, \ldots, \boldsymbol{\pi}^{(H)}).
\end{eqnarray*}}}Recalling results in \citet{dun_2009} a sufficient condition for the above integral to be strictly positive is that $\Pi_y\{{p}_y: \sum_{y=1}^2 | p_{\mathcal{Y}}(y)-p_{\mathcal{Y}}^0(y)|<\epsilon_y\}>0$,  $\Pi_{\pi}\{\boldsymbol{\pi}^{(1)}, \ldots, \boldsymbol{\pi}^{(H)}: \sum_{h=1}^H \sum_{l=1}^{V(V-1)/2} | \pi^{(h)}_l-\pi^{0(h)}_l |<\epsilon_{\pi}\}>0$ and $\Pi_{\nu}\{\boldsymbol{\nu}_1, \boldsymbol{\nu}_2: \sum_{y=1}^2 \sum_{h=1}^H | \nu_{hy}-\nu^0_{hy}|<\epsilon_{\nu}\}>0$ for every $\epsilon_{\pi}>0$, $\epsilon_y>0$ and $\epsilon_{\nu}>0$. The large support for ${p}_{\mathcal{Y}}$ is directly guaranteed from the Beta prior. Similarly, according to Theorem 3.1 and Lemma 3.2 in \citet{dur_2014}, the same hold for the joint prior over the sequence of component-specific edge probability vectors $\boldsymbol{\pi}^{(h)}, h=1, \ldots, H$ induced by priors $\Pi_Z$, $\Pi_X$ and $\Pi_{\lambda}$ in factorization (2.9). Finally, marginalizing out the testing indicator $T$ and recalling our prior specification for the mixing probabilities in (3.1) a lower bound for $\Pi_{\nu}\{\boldsymbol{\nu}_1, \boldsymbol{\nu}_2: \sum_{y=1}^2 \sum_{h=1}^H | \nu_{hy}-\nu^0_{hy}|<\epsilon_{\nu}\}$ is
\begin{eqnarray*}
\mbox{pr}(H_0)\Pi_{\upsilon} \{ \boldsymbol{\upsilon}: \sum_{y=1}^2 \sum_{h=1}^H | \upsilon_{h}-\nu^0_{hy}|<\epsilon_{\nu}\}+\mbox{pr}(H_1) \prod_{y=1}^{2} \Pi_{\upsilon_y} \{ \boldsymbol{\upsilon}_y :\sum_{h=1}^H | \upsilon_{hy}-\nu^0_{hy}|<\epsilon_{\nu}/2\}.
\label{app_1}
\end{eqnarray*}
If the true model is generated under independence, the above equation reduces to 
\begin{eqnarray*}
\mbox{pr}(H_0)\Pi_{\upsilon} \{ \boldsymbol{\upsilon}: \sum_{h=1}^H | \upsilon_{h}-\nu^0_{h}|<\epsilon_{\nu}/2\}+\mbox{pr}(H_1) \prod_{y=1}^{2} \Pi_{\upsilon_y} \{ \boldsymbol{\upsilon}_y :\sum_{h=1}^H | \upsilon_{hy}-\nu^0_{h}|<\epsilon_{\nu}/2\},
\end{eqnarray*}
with the Dirichlet priors for $\boldsymbol{\upsilon}$, $\boldsymbol{\upsilon}_1$ and $\boldsymbol{\upsilon}_2$ ensuring the positivity of both terms. When instead $\nu^0_{h1} \neq \nu^0_{h2}$ for some $h=1, \ldots, H$, the inequality $\mbox{pr}(H_0)\Pi_{\upsilon} \{ \boldsymbol{\upsilon}: \sum_{y=1}^2 \sum_{h=1}^H | \upsilon_{h}-\nu^0_{hy}|<\epsilon_{\nu}\}>0$ is not guaranteed, but $\mbox{pr}(H_1) \prod_{y=1}^{2} \Pi_{\upsilon_y} \{ \boldsymbol{\upsilon}_y :\sum_{h=1}^H | \upsilon_{hy}-\nu^0_{hy}|<\epsilon_{\nu}/2\}$ remains strictly positive for every $\epsilon_{\nu}$ under the independent Dirichlet priors for the quantities $\boldsymbol{\upsilon}_1$ and $\boldsymbol{\upsilon}_2$, proving the Proposition.
\end{proof}

\vspace{-20pt}

\bibliographystyle{ba}
\bibliography{example}

\begin{acknowledgement}
This work was partially funded by the grant CPDA154381/15 of the University of Padova, Italy, and by the Office of Naval Research grant N00014-14-1-0245, United States. The authors would like to thank Rex E. Jung and Sephira G. Ryman for the brain connectivity data and creativity scores funded by the John Templeton Foundation (Grant 22156) entitled ``The Neuroscience of Scientific Creativity.'' The authors are also grateful to William Gray Roncal and Joshua T. Vogelstein for help in accessing the connectome data. We finally thank the Editor, the Associate Editor and the referees for the valuable comments on a first version of the article.
\end{acknowledgement}

\end{document}